\documentclass [12pt,a4paper]{article}

\usepackage{graphicx}
\usepackage{amsfonts}    \usepackage{amssymb}
\usepackage{amsmath}
\newcommand{\D}[2]{\frac{\partial #2}{\partial #1}}
\newcommand{\DD}[2]{\frac{\partial^2 #2}{\partial #1^2}}

\renewcommand{\vec}[1]{\mbox{\boldmath$#1$}}

\newcommand{\eps}{\varepsilon}

\begin{document}

\title{ Thermo-Mechanical Wave Propagation In Shape Memory Alloy Rod With
Phase Transformations}
\author{L. X. Wang$^1$
\thanks{Corresponding author: Tel: 45+6550 1686. E-mail:
wanglinxiang@mci.sdu.dk }
, and  R. V. N. Melnik$^{1,2}$  \\
 $^1$ University of Southern Denmark,\\
Grundtvigs Alle 150  \\
 Sonderborg, DK-6400, Denmark \\
$^2$ Wilfrid Laurier University \\
75 University Avenue West,  \\
Waterloo, Ontario, Canada, N2L 3C5 \\  }
\date{}

\maketitle


\begin{abstract} Many new applications of ferroelastic materials require
  a better understanding of their dynamics that often involve phase
  transformations. In such cases, an important prerequisite is the
  understanding of  wave
  propagation caused by pulse-like
loadings. In the present study, a mathematical model  is developed
 to analyze the wave propagation process in shape memory alloy rods. The first order martensite
transformations and associated thermo-mechanical coupling effects are accounted
for by employing the modified Ginzburg-Landau-Devonshire theory.
The Landau-type free energy  function is employed to characterize different phases,
while a Ginzburg term is introduced to account for energy contributions from phase boundaries.
The effect of internal friction is represented by a Rayleigh dissipation term.
The resulted nonlinear system of PDEs is reduced to a differential-algebraic system, and
Chebyshev's collocation method is employed together with the backward differentiation method.
A series of numerical experiments are performed. Wave propagations caused
by impact loadings are  analyzed for different initial temperatures. It is demonstrated that
coupled waves will be induced in the material. Such waves will be dissipated and dispersed
during the propagation process, and phase transformations in the material will complicate
 their propagation patterns. Finally, the influence of internal friction and capillary effects
 on the process of wave propagation is analyzed numerically.


\end{abstract}


\section{INTRODUCTION}

In the past decades, different aspects of Shape Memory Alloys  (SMA)  have been
investigated intensively by mathematicians, physicists, and engineers  \cite{Birman1997}.
 This interest in a larger scientific community
 is due to SMA unique properties of being able to convert thermal energy into mechanical and
vice versa.  These properties are promising for many applications of SMAs, including mechanical and
control engineering, biomedicine, communication, robotics to name just a few \cite{Birman1997}.
Motivated by application developments of advance composite materials involving SMAs,
 nonlinear wave propagations in these materials have
been investigated as a stepping ground for the prediction and understanding of dynamic
 response of the composite under various dynamic loadings \cite{Abeyaratne1994,Berezovski2005,Falk1984}.

Compared to wave propagations in conventional solid materials, the impact induced wave
propagations in  materials such as SMAs requires delicate treatments as additional
 difficulties arise due to phase transformations \cite{Abeyaratne1994,Berezovski2005,Falk1984}.
In general, impact loadings of these materials will cause nonlinear
thermo-mechanical waves which are similar to those of other
thermo-elastic materials under impact loadings. The difference of
wave propagation in conventional solids and those in the
ferroelastic materials such as SMAs is that the first order
martensitic transformation may be induced in the latter case. The
transformation is reversible, and its native nonlinearity and
hysteresis will have a substantial influence on the wave propagation
 and will make the wave propagation patterns
 more complicated \cite{Abeyaratne1994,Berezovski2005,Falk1984}.

The first step to the modelling of impact induced wave propagations
and phase transformations is a sound constitutive theory upon which
the entire model can be built \cite{Abeyaratne1994,Falk1984}.
Various constitutive models have been  proposed
 on mesoscale and microscale to capture the phase boundary movement induced by
  the dynamical loadings  \cite{Abeyaratne2000,Dai2006}. Example of such a
 constitutive model can be found in Ref. \cite{Abeyaratne1994,Abeyaratne2000} where
 an one-dimensional model for the modelling of shock
wave propagations with phase transformation was constructed on the basis of a non-convex
Helmholtz free energy function. Under this approach the entire structure was split into
different domains due to the phase transformation and the movement of boundaries
between the
 domains was modelled using the ``jump conditions". This approach is suitable for
microscopic problems, while for many engineering applications a
model is required at macroscale. In Ref.
\cite{Chen2000,Lagoudas2003}, the dynamic behaviour of
 phase boundaries was modelled using a thermo-mechanical coupling approach.
 The model was based on a linearized
constitutive theory, and hence its application potential was
inherently limited.

For many engineering applications, the dynamic response of SMA
materials caused by impact loadings need to be better understood at
macroscale for design and control of SMA-based devices. For this
purpose, displacement and temperature evolutions in the material are
normally sought. Models at mesoscale may not be sufficient for this
purpose as another model needs to be constructed to link macroscale
properties and mesoscale domain structures.
 Another aspect of  modelling the dynamics of such materials as SMAs
 under impact loadings is the thermo-mechanical coupling effects. In most of the existing
investigations, the thermal dynamics are either neglected \cite{Abeyaratne1994,
Berezovski2005,Dai2006},  or modelled separately from the mechanical dynamics
\cite{Chen2000,Lagoudas2003}.  However, the physics-based models should account for the
intrinsic coupling of thermal and mechanical fields in SMAs. When the SMAs are used for
damping purposes or for other purposes where the conversion of energy between the thermal and
mechanical fields is essential, the coupling effects are expected to be particularly
important, and the constitutive theory should  be constructed by taking into account both
fields simultaneously.

In this paper, the nonlinear thermo-mechanical wave propagations in
SMA rods induced by impact loadings are modelled and analyzed at
macroscale.  To capture the thermo-mechanical coupling and nonlinear
nature of the phase transformations,  the Ginzburg-Landau-Devonshire
theory is applied for the modelling of the nonlinear dynamics. The
governing equations for the mechanical field are obtained by
minimizing the mechanical energy, while those for the thermal field
are obtained by using the conservation law of internal energy.  The
intrinsic coupling of the two fields is built-in into the model by
including both fields in the potential energy functional. In the
following sections, a mathematical model describing SMA dynamics is
developed based on a system of coupled partial differential
equations which is re-cast in the form of differential algebraic
equations, and Chebyshev' collocation method
 is employed together with the backward differentiation formula to
  integrate the resulting system. Nonlinear wave propagation patterns caused by an impact stress
loading at one end of the rod are simulated with different initial
temperatures and computational parameters. Finally, the influence of
phase transformation on the wave propagations is analyzed
numerically, along with the influence of other effects such as
internal friction and capillary effects.


\section{THE INITIAL-BOUNDARY VALUE PROBLEM}

\begin{figure}    \begin{center}
     \includegraphics[scale=0.35]{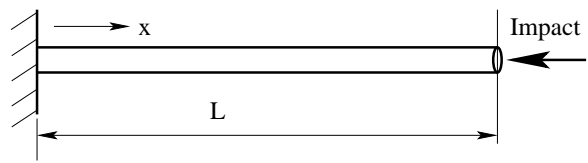}
     \caption{Shape memory alloy rod under impact loadings}
     \label{RodShock}
     \end{center}  \end{figure}

We restrict our analysis to one-dimensional cases as sketched in Figure (\ref{RodShock}). The
SMA rod under consideration occupies an interval $[0, L]$, and  is subjected to an impact
loading from the right end $x=L$, while the other end $x=0$ is fixed. The rod is thermally
insulated at both ends so there is no heat loss (gain) to (from) the ambient environment.
Under external loadings, a material point $x$ in the SMA rod will be moved to a new position
$x+u(x,t)$ due to deformation, where $u(x,t)$ is the longitudinal displacement at time $t$.
Function $u(x,t)$ is assumed to be continuous at any time $t$ and position $x$ based on the
continuity of the rod at  macroscale. The stress $\sigma$ is related to the deformation $\eps$
 by $\sigma(x,t) = \mathcal N (\eps(x,t))$ where  $\eps(x,t)= \partial u(x,t)/ \partial x$
 is the strain.

By now, it is well understood that the first order phase
transformation in SMAs holds the key to unique properties of the
material, such as the shape memory and pseudo-elastic effects.
Therefore, it is expected that the adequate mathematical model for
the dynamics of SMAs at macroscale should be able of capturing the
first order martensite phase transformations. On the other hand, the
intrinsic coupling of mechanical and thermal fields should also be
captured by the model. To satisfy these requirements, the
mathematical model based on the modified Ginzburg-Landau-Devonshire
theory has been established
 \cite{Bubner1996,Bubner2000,Falk1984}:
    \begin{eqnarray} \label{GoverningEq}
    \rho \ddot u &=& \D{x}{}\left ( k_1(\theta -\theta_1) \eps  + k_2 \eps^3  + k_3 \eps^5 \right )
      + \nu \D{t}{}\DD{x}{u}  -  k_g \frac{\partial ^4u}{\partial x^4} + f, \\
    c_{v}\D{t}{\theta}&=& k\DD{x}{\theta}+k_{1}\theta \eps \D{t}{\eps} +
        + \nu \left ( \D{t}{\eps} \right)^2 + g ,
    \end{eqnarray}
 where $k_{1}$, $k_{2}, k_{3}$ and  $k_g$ are material-specific constants, $\nu$
 internal friction coefficient,  $\rho$ the density of the material,  $\theta_1$ the reference
  temperature,  $c_v$ the specific heat capacitance,  $k$ heat conductance, and $f$ and
  $g$ mechanical and thermal loadings, respectively.

It is essential that the above model  is constructed on the basis of
the potential energy function $\mathcal F$, which a non-convex
function
 of the chosen \emph{order parameters} and temperature $\theta$, at mesoscale,  according to
Ginzburg-Landau-Devonshire theory. It is a sum of local energy
function $(\mathcal F_l)$ and non-local energy function ($\mathcal
F_g$).  For the current one-dimensional  problem,
 the strain $\eps(x,t)$ is  chosen as the order parameter,  and the local free energy density
 can  be constructed as the Landau free energy density $\mathcal F_l(\theta, \eps)$, while
 the non-local part can be formulated as a Ginzburg term $\mathcal F_g( \nabla \eps)$:
 \cite{Bales1991,Chaplygin2004,Falk1984}:
    \begin{eqnarray} \label{Potentials}
    \mathcal F(\theta, \eps) &=&  \mathcal F_l(\theta, \eps)  +     \mathcal F_g( \nabla \eps) ,
                 \nonumber  \\
    \mathcal F_l(\theta, \eps) &=&  \frac{k_1(\theta -\theta_1)}{2} \eps^2  +
     \frac{k_2}{4} \eps^4  + \frac{k_3}{6} \eps^6, \\
    \mathcal F_g (\nabla \eps) &=& k_g (\D{x}{\eps})^2, \nonumber
    \end{eqnarray}
The local minima of the local term (Landau free energy function) are introduced to characterize
martensite variants, while the non-local term (Ginzburg term) above accounts for
inhomogeneous strain field, which represents energy contributions from domain walls and
other boundaries among different phases. It will be translated into capillary effects at
macroscale.

In order to account for the internal friction, accompanying wave
propagations and phase transformations, which will be translated
into viscous effects at macroscale, a Rayleigh dissipation term is
also included in the above model by using the following term
 \cite{Abeyaratne2000}:
    \begin{eqnarray}
         \mathcal F_R = \frac{1}{2}\nu (\D{t}{\eps})^2.
    \end{eqnarray}

The constitutive relations for the material in the above model at
macroscale can be obtained by using the thermodynamic equilibrium
conditions:
    \begin{eqnarray}
     \sigma = \D{\eps}{H},    \quad  e= \mathcal H - \theta  \D{\theta}{\mathcal H},
    \end{eqnarray}
where $\mathcal H(\theta, \eps)= \mathcal F  - c_v \theta \ln \theta$ is the Helmholtz free
energy function.  The mechanical and thermal fields are intrinsically coupled since the
internal energy $e$ is associated with the same potential energy as above, and the
governing equations for the thermal field are formulated using the conservation law for the
internal energy \cite{Bubner1996,Falk1984}.

The difficulties associated with the above model are understood
better by analyzing the profiles of the non-convex potential energy
$F_l$, its temperature dependence, and the non-convex constitutive
curves,
 as sketched in Figure~(\ref{GinzburgLandau}).  At low temperature $\theta=201$ K, as sketched in
the top row,  there are two local minima in the potential function,
which are introduced to characterize martensite plus and minus,
respectively (in 1D cases). The stress-strain relation of the
material under dynamical loadings will not follow the sketched
constitutive relations exactly. Instead, there will be jumps from
point ($A$ to $B$) or vice versa. Such jumps
 are associated with the transition from martensite plus to minus or vice versa.
 This is the origin of mechanical hysteresis which will dissipate mechanical energy quickly by
 converting it into thermal form due to the thermo-mechanical coupling.
 The amount of energy converted into  thermal form can be estimated by using the area enclosed
 under the  hysteresis loop, as marked by the dashed lines in the stress-strain plot on the left.
When the SMA rod temperature is intermediate ($\theta=240$ K), there
are still hysteresis loops, and the jump phenomena become more
complicated, since there are three local minima present in the
potential energy function, which means that austenite and two
 martensite variants may co-exist in the material. The dissipation of mechanical energy
 will be slower in this case
since the hysteresis loop become smaller.  At high temperature
($\theta=310$ K), it is shown that there are no jump phenomena any
longer, therefore no hysteresis, because there is only one local
 minimum. In this case, there will be only austenite present and the dynamics become
 fairly simple.

In the current paper, the mechanical loading is implemented in terms
of impact stress at one of the rod ends.  Hence, it is convenient to
keep the constitutive relation
 as an extra equation for the model and consider the stress as a dependent variable.
 This representation will make the treatment of boundary conditions much easier for the
 current discussion. The resulting system of Differential Algebraic Equations (DAE) can
 be written as follows \cite{Melnik2000}:
\begin{equation} \label{SMAEq}
  \begin{array}{l}
 \displaystyle    \D{t}{u} = v, \quad
 \rho\D{t}{v} =  \D {x}{\sigma}  + \nu \D{x}{} \D{x}{v}
   -  k_g  \D{x}{}\DD{x}{\eps},  \\
 \displaystyle
c_{v}\frac{\partial\theta}{\partial
t}=k\frac{\partial^{2}\theta}{\partial x^{2}}+k_{1}\theta \eps
\D{t}{\eps}, \\
 \displaystyle
\sigma =  k_{1}(\theta-\theta_{1}) \eps  +  k_{2}\eps^{3}+k_{3}\eps^{5},
\end{array}
\end{equation}
where $v$ is the velocity. The mechanical and thermal loadings, $f$
and $g$, are all set to zero, so that only boundary loadings will be
taken into account in the current investigation.

In order to investigate the thermo-mechanical wave propagations in
the SMA rod, the following boundary conditions are employed for the
mechanical and thermal fields similarly as in
Ref.\cite{Bubner1996,Wang2006} :
    \begin{eqnarray}
     \left. \D{x}{\theta} \right |_{x=0}  =0,  \quad \left. \D{x}{\theta}\right |_{x=L}  =0,
     \nonumber   \\
          u(0, t)   = 0,   \quad \sigma(L, t) = S(t), \\
       \left. \DD{x}{u}\right |_{x=0}   = 0, \quad    \left. \DD{x}{u} \right |_{x=L}=0.
    \end{eqnarray}
where $S(t)$ is a given function describing the stress impact profile.


\section{Wave Propagations}

Since mechanical responses caused by external loadings in most materials are
normally much faster than thermal ones,  which are also more interesting for numerical
 investigations and many applications, the emphasis of the current discussion is
put on mechanical waves caused by mechanical loadings.

\subsection{Temperature Dependence}

For the analysis of elastic waves in the SMA rod, Equation~(\ref{GoverningEq}) is firstly
linearized at the point $(\sigma_L, \eps_L)$ where $\sigma_L=0$:
    \begin{eqnarray}\label{WaveLinear}
  \rho \ddot u &=& \D{x}{}k_L \eps + \nu \D{t}{}\D{x}{\eps}  -
   k_g \frac{\partial ^4u}{\partial x^4},
    \end{eqnarray}
where the external mechanical loading is dropped out. $k_L$ is the stiffness constant for the
linearized system, which is temperature dependent.

When the SMA is at high temperature ($\theta=310$ K), only austenite is stable and there is no
 phase transformation, it can be easily calculated that
$\eps_L=0$ so $k_L$ can be simply formulated as: $ k_L  = k_1(\theta -\theta_1)$.

\begin{figure}   \begin{center}
    \includegraphics[width=7.6cm, height=6cm]{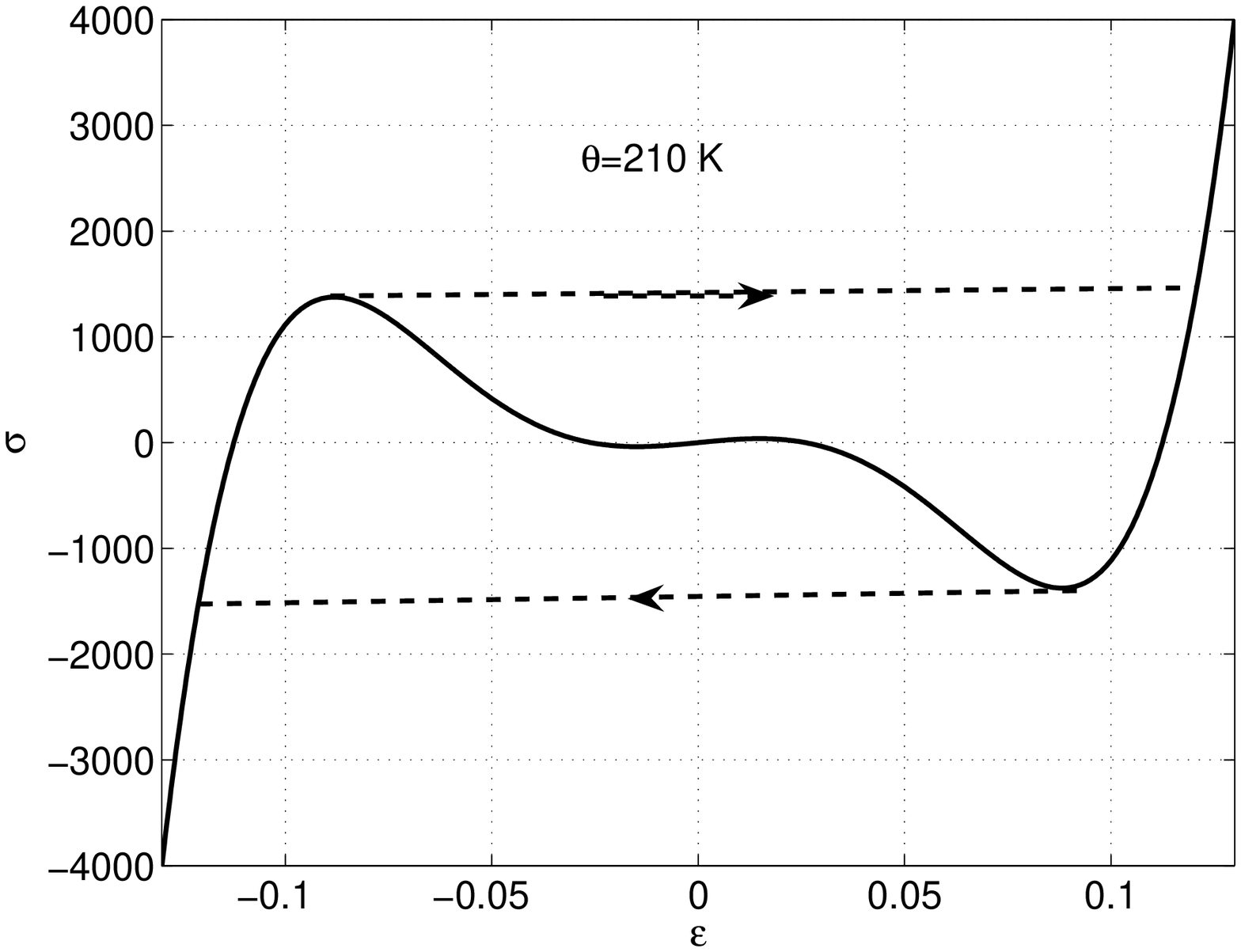}
    \includegraphics[width=7.6cm, height=6cm]{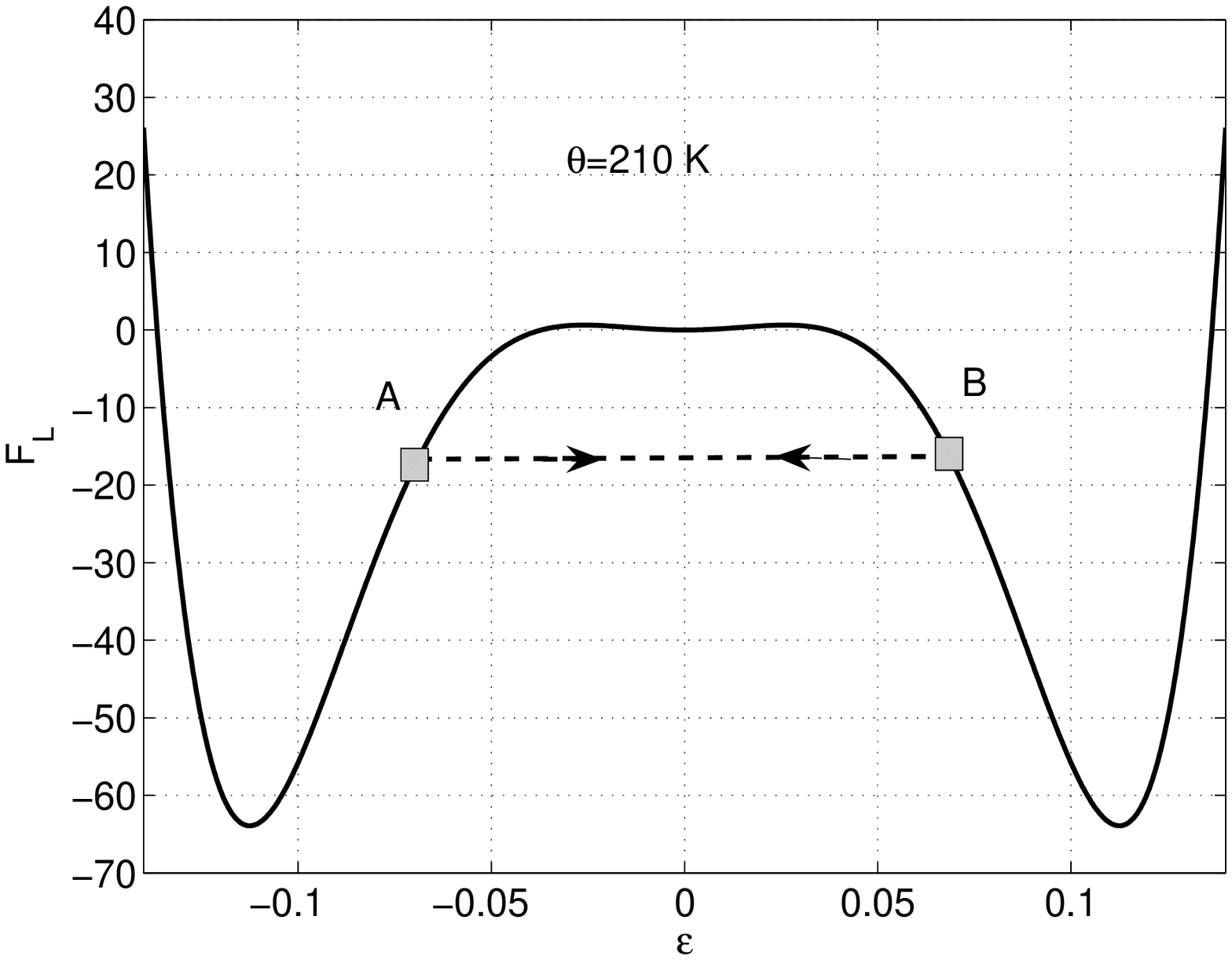} \\
    \includegraphics[width=7.6cm, height=6cm]{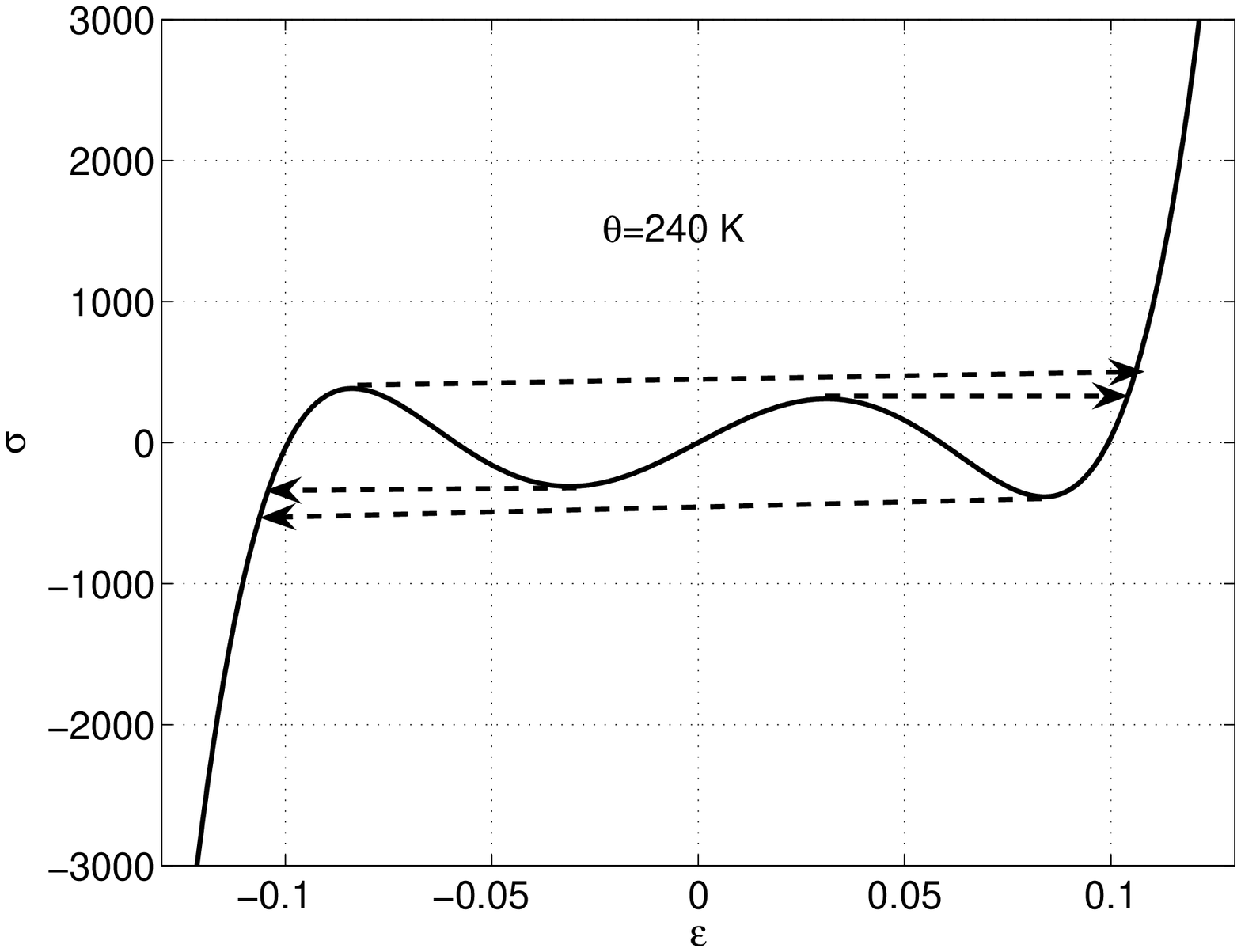}
    \includegraphics[width=7.6cm, height=6cm]{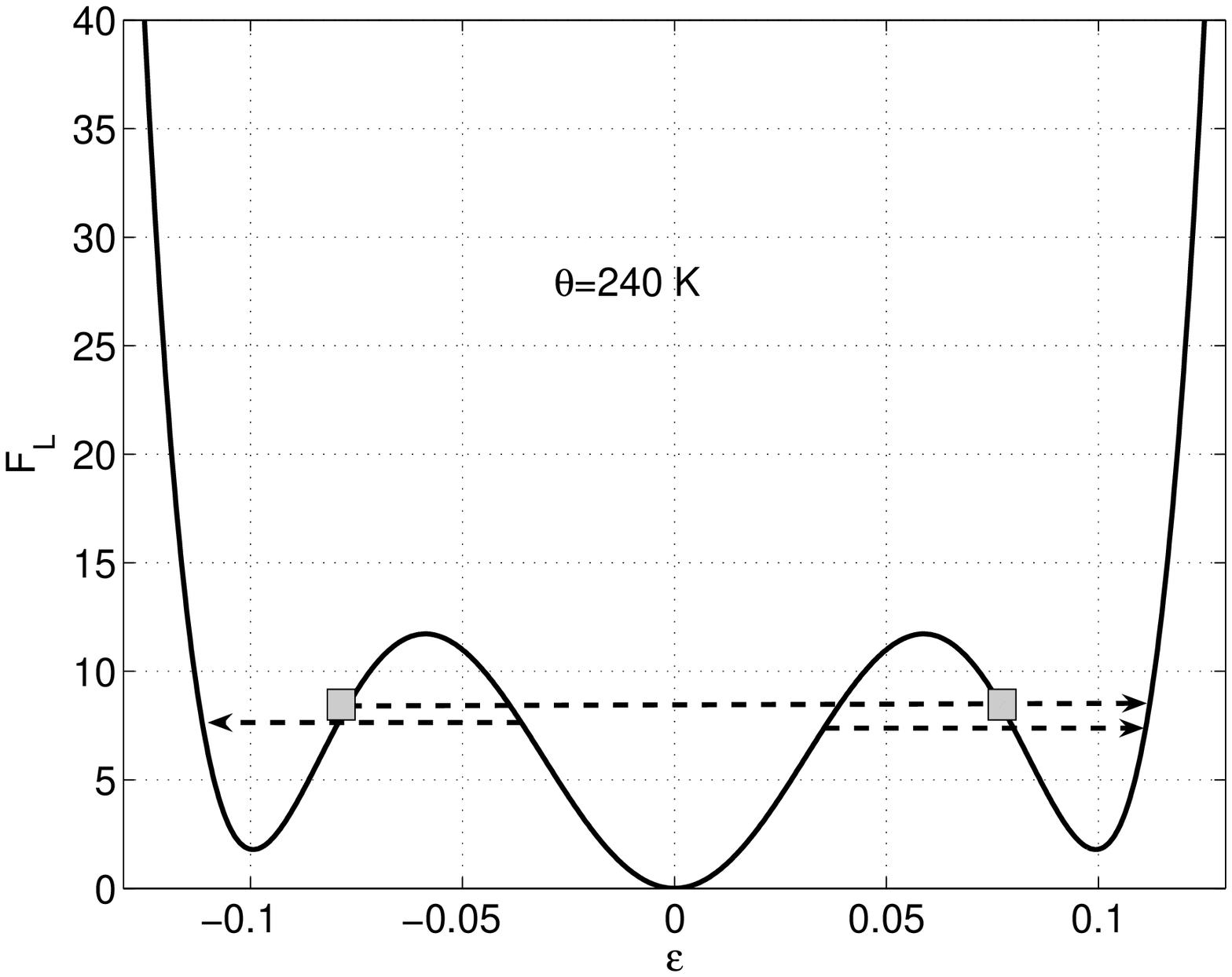} \\
    \includegraphics[width=7.6cm, height=6cm]{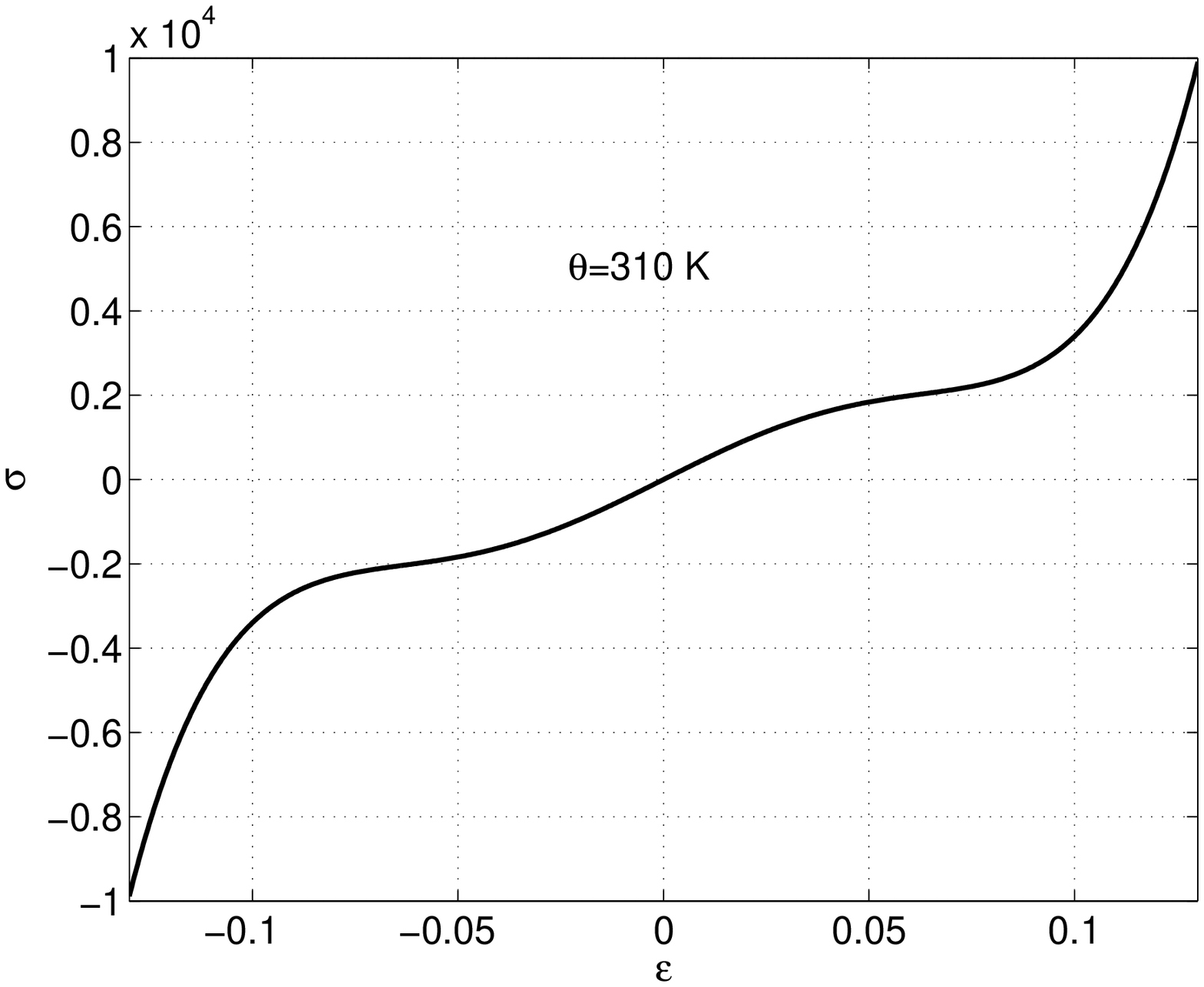}
     \includegraphics[width=7.6cm, height=6cm]{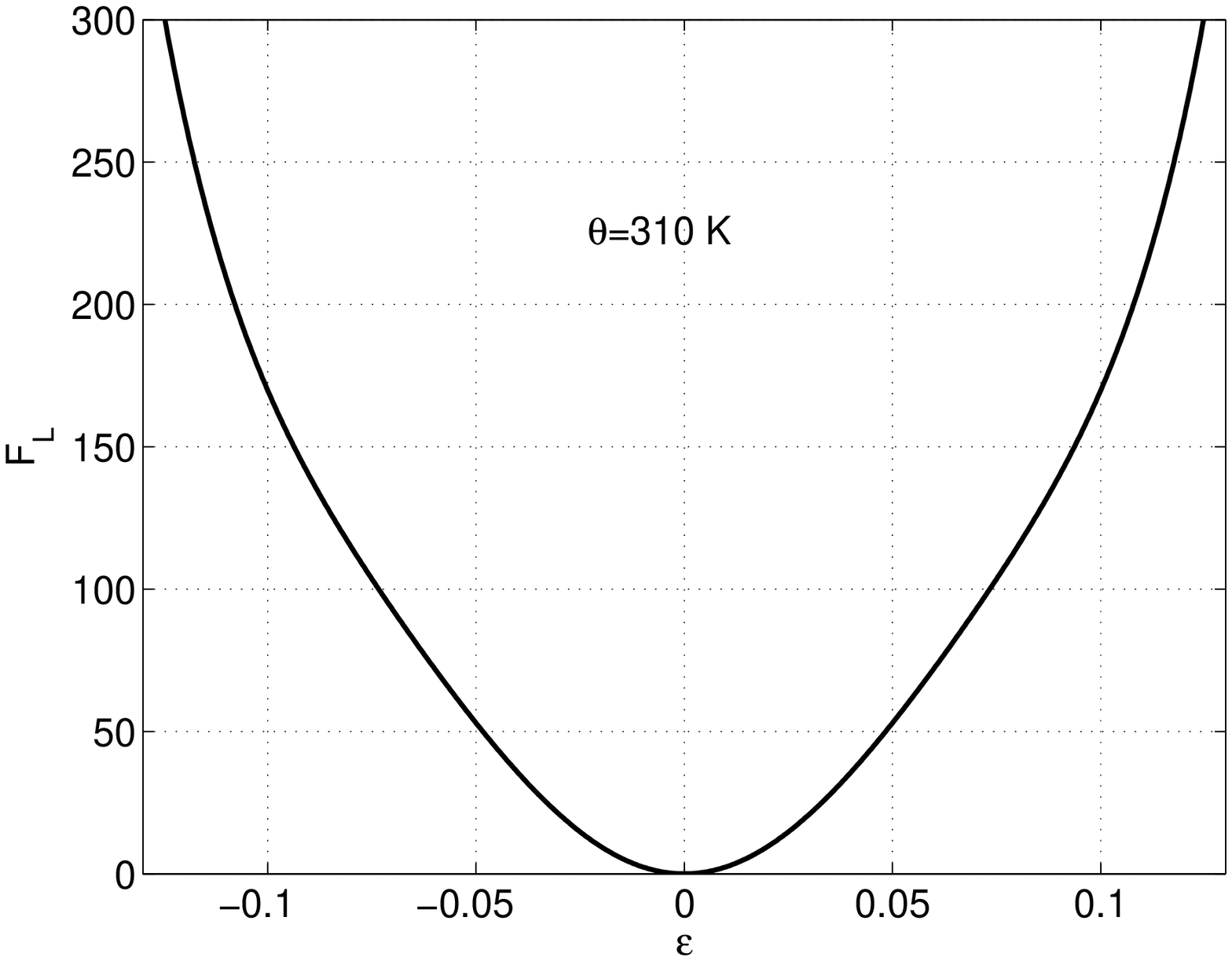}
     \caption{Constitutive relations (left colum) for a shape memory alloy and the associated
     potential energy density (right colum), at various temperatures. (top) $\theta = 210$ K.
      (middle) $\theta = 240$ K. (bottom) $\theta = 310$ K. }
     \label{GinzburgLandau}
      \end{center}  \end{figure}

At lower temperature ($\theta=210$ K), the stress strain relation is
not a monotone curve any longer, and the linearization has to be
carried out by using at least three points, as indicated by the plot
in the top row of Figure~(\ref{GinzburgLandau}) (the three
intersections between the horizontal axis with the $\sigma-\eps$
curve). The central one is $\eps_L=0$, which is not a stable
equilibrium state of the system, and is not interesting for the
analysis. The other two intersections can be easily calculated using
the following condition:
    \begin{eqnarray}
     k_1(\theta -\theta_1)  +  k_2 \eps^2  + k_3 \eps^4= 0,
    \end{eqnarray}
which gives the following formulation (using parameter values given in section 5):
    \begin{eqnarray}  \label{EquiPoint}
    \eps_L^2= \frac{-k_2 \pm \sqrt{ k_2^2-4 k_1k_3(\theta-\theta_1) } }{2 k_3},
    \quad \eps_L = \pm 0.115.
    \end{eqnarray}
These two values are associated with strains for martensite plus ($\eps_L=0.115$)
and minus ($\eps_L=-0.115$), respectively.
The linearized coefficient then can be calculated as follow:
    \begin{eqnarray}
     k_L =  k_1(\theta -\theta_1)  +  k_2 \eps_L^2  + k_3 \eps_L^4,
    \end{eqnarray}
which gives $k_L$ the same value at the two points  with different $\eps_L$ values,
 due to the symmetry property. The above analysis indicates that the
 linearized wave motion in the material at martensite plus state will be the same
 as those at martensite minus state.

 For the cases where the SMA  rod temperature is intermediate, the dependence
  of $k_L$ on temperature is more
 complicated since both austenite and martensite might occur and there might be $4$
 values for $\eps_L$ to satisfy $\sigma_L=0$, as indicated in the middle row in
Figure~(\ref{GinzburgLandau}). Although the symmetry property still
exists for martensitic variants, wave motions are different between
austenite and martensite states.

Let us consider the following wave solution to the linearized wave
motion given by Equation~(\ref{WaveLinear}):
    \begin{eqnarray}
     u = u(x- V t) = u(z),  \quad z = x- V t,
    \end{eqnarray}
where $V$ is a constant stands for wave velocity.  By substitution,  the following relation
can be easily obtained:
\begin{eqnarray} \label{WaveSimplified}
     k_g\DD{x}{\eps} =  \left( k_L -\rho V^2  \right) \eps  - \nu V  \D{x}{\eps},
\end{eqnarray}
where $ K_L  -\rho V^2$ can be positive or negative.  The problem is now formulated
as an ordinary differential equation and its general solution can be written as:
    \begin{eqnarray}
           \eps(z) = (C_1 + C_2 z) e^{z \sqrt{ (k_L -\rho V^2 )/k_g}},
    \end{eqnarray}
where $C_1, C_2$ are coefficients to be determined by boundary
conditions. The viscous term is temporarily ignored.

 If the wave velocity $V$ is less than the velocity of sound in the material
 $V_s =\left ( k_L/\rho \right)^2$, then  $ K_L  -\rho V^2 >0$, and there is no limited solution
  exists. It means that no waves can propagate in the material with velocity
   $V$ smaller than the velocity of sound in  the material. Since the velocity of sound of the material
   is temperature dependent, so the allowed speed for waves to propagate in the SMA
   rod varies along with the variation of its temperature.

When the viscous term is also taken into account, then the wave propagation will
always be accompanied by dissipation effects, which can be characterized by the
following exponential function:
    \begin{eqnarray}
        | \eps(z)| = e^{-\xi \omega z}, \quad  \xi =\frac{\nu V}{2\sqrt{(\rho V^2 -k_L) k_g}},
        \quad   \omega = \sqrt{\frac{(\rho V^2 -k_L)}{k_g}},
    \end{eqnarray}
where the initial amplitude of the considered waves are assume to be
$1$. The dissipation effects can be estimated by the exponential
coefficient $\xi \omega =  \frac{\nu V}{2k_g}$. For larger $V$,
faster dissipation will be induced. The dissipation effects are
independent of the material temperature.

\subsection{Effects of Ginzburg's Term}

In the wave equation, the term $k_g \frac{\partial^4 u}{\partial
x^4}$ is resulted from the interfacial energy contribution to the
potential energy function. It is also called Ginzburg's term,  and
accounts for the capillary effects
 \cite{Bubner1996,Bubner2000,Falk1984,Falk1987}.
It is easy to see that this term is related to dispersion of wave propagations. For the
analysis,  the following solution to the linearized wave equation is considered:
\begin{eqnarray}
    u =  \sin\frac{2\pi}{\lambda }(x - Vt),
    \end{eqnarray}
where $\lambda$ is the wave length. By substitution, the following relationship involves
the wave speed $V$ can be obtained if the viscous effects are ignored:
    \begin{eqnarray}  \label{WaveDisper}
    \rho V^2 = k_L + \frac{4\pi k_g}{\lambda^2},
    \end{eqnarray}
which indicates that the wave propagation speeds are increased due
to the contribution of non-local potential energy (capillary
effects). The dispersion effects caused by non-local contributions
are stronger for those waves with smaller wave lengthes.  If the SMA
rod is at lower temperature, the linearized stiffness constant $k_L$
will be smaller, which will make the last term in
Equation~(\ref{WaveDisper}) more pronounced.


\section{NUMERICAL METHODOLOGY}

As mentioned in the previous section,  the development of the
numerical methodology for simulation of the wave propagation based
on the above mathematical model is not a trivial task. In
particular, given that both dispersion and dissipation of wave
propagations are present in the physics of the problem, the
numerical algorithm for the problem has to be able to take care of
both dissipation and dispersion numerically, and  the accuracy of
the algorithm will be affected by their treatment. In the present
paper, a multi-domain decomposition method combined with the
Chebyshev collocation methodology is the method of choice in
addressing the above issues. The compromise made here is that a
spectral method is employed  to take advantage of its better
convergence property, while the domain decomposition method is
chosen for the purpose to reduce the order of basis functions for
the spectral method when the total node number is large.

\subsection{Chebyshev's Collocation Method}

For the Chebyshev pseudo-spectral approximation, a set of
Chebyshev points $\{x_i\}$ are chosen along the length direction as follows:

\begin{equation}
  \label{num-eq4}
  x_i  = L\left(1-\cos(\frac{\pi i}{N})\right) /2, \quad i=1,2,\dots,N.
\end{equation}

Using these nodes, $u, v, \theta$, and  $\sigma$ distributions in the rod can
be expressed in terms of the following linear approximation:

\begin{equation}
  \label{num-eq2}   f(x) = \sum_{i=1}^{N} f_i \phi_i(x),
\end{equation}
where $f(x)$ stands for any of  $u, v, \theta$, or  $\sigma$,
 and $f_i$ is the function value at $x_i$.  $\phi_i(x)$ is the $i^{th}$ interpolating
polynomial which has the following property:
   \begin{equation}
     \phi_i(x_j) = \left \{
     \begin{array}{ll}
       1,  & i=j, \\  0, & i\neq j.
     \end{array}  \right.
   \end{equation}

It is easy to see that the well-known Lagrange interpolants satisfy
the interpolating requirements.  Having obtained $f(x)$
approximately, the derivative $\partial f(x)/ \partial x$ can be
easily obtained by taking the derivative of the basis functions
$\phi_i(x)$ with respect to $x$:
 \begin{equation}
   \label{num-eq3}
  \frac{\partial f}{\partial x} = \sum_{i=1}^{N} f_i
  \frac {\partial \phi_i(x)}{\partial x},
\end{equation}
and similarly for the higher order derivatives. All these approximations can be formulated in
the matrix form, for the convenience of programming.

\subsection{Multi-Domain Decomposition}

It is known that the  spectral methods are able to give a higher
accuracy with the same number of discretization nodes, compared to
finite difference methods or finite element methods. On the other
hand, when the solution to the problem does not have higher-order
derivatives, the spectral methods may lead to artificial
oscillations due to the Gibbs phenomenon. This may be expected for
the current problem when the impact induced wave propagation is
analyzed numerically. To avoid this, a multi-domain decomposition
method is employed.

\begin{figure}    \begin{center}
     \includegraphics[scale=0.7]{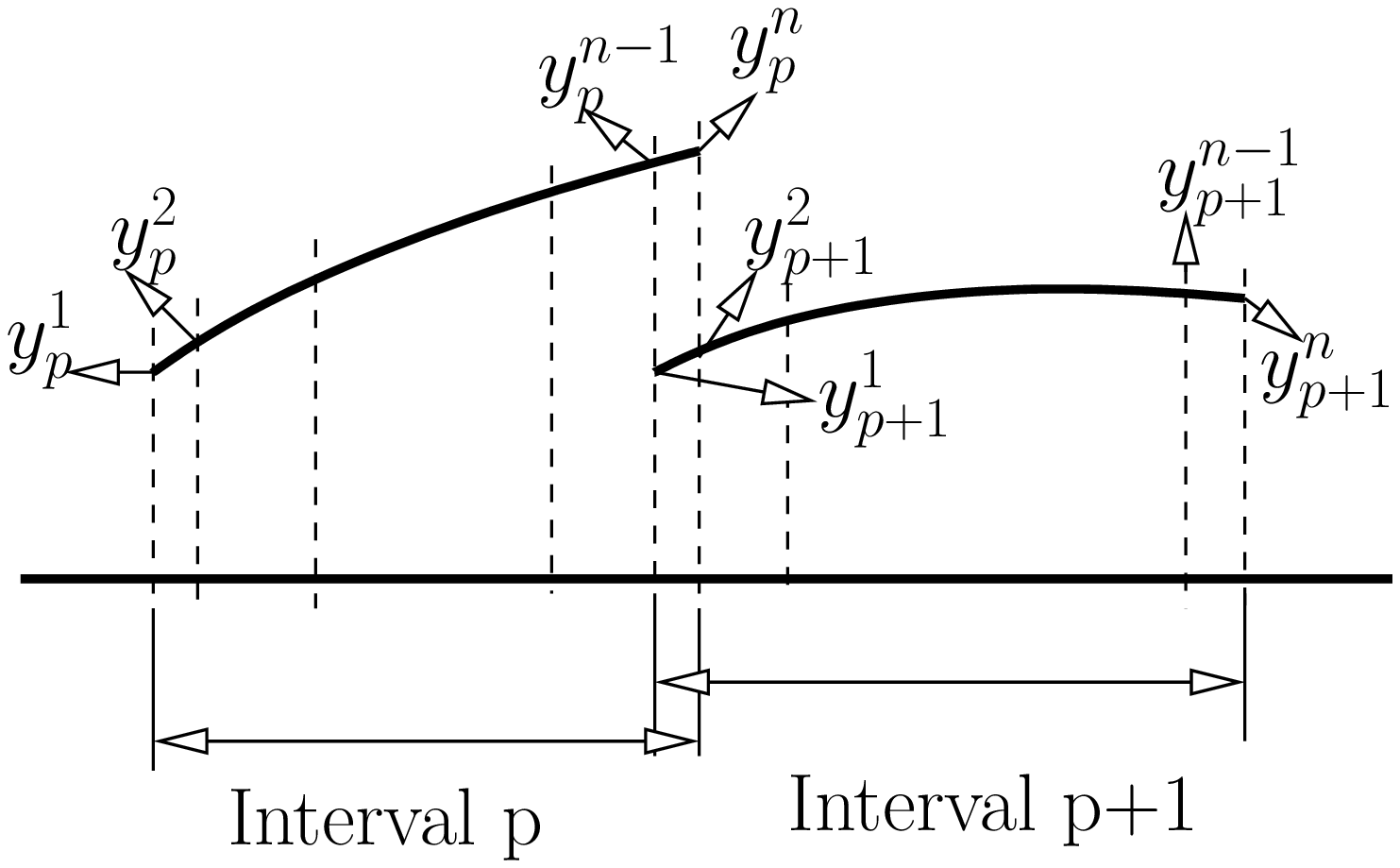}
     \caption{Sketch of domain decomposition and discretization}
     \label{DomainDec}
      \end{center}  \end{figure}

The entire computational domain $\mathcal D =  [0, L]$ is evenly decomposed into $P$
intervals (subdomains), with an overlap region between each
 pair of consecutive intervals, as sketched in Figure (\ref{DomainDec}):
    \begin{equation}
    \mathcal D = \bigcup_{p=1}^{p=P} D_p,
    \end{equation}
where the number of subdomains $P$ is chosen according to the specific problem under
consideration. In each interval, the Chebyshev collocation method discussed in the previous
section is employed to approximate the solution and its derivatives.

The coupling between each pair of consecutive intervals can be implemented
by setting the following requirements:
    \begin{eqnarray}
         y_p^n = y_{p+1}^2, \quad      y_p^{n-1} = y_{p+1}^1,
    \end{eqnarray}
where the subscript $p$ stands for the interval number, while the superscript $n$  stands for
the node number in each interval.  Variable $y_p^n$ is the function value at point $x_p^n$
(the $n$th node in the $p$th interval), which could be any of the dependent variables we are
solving for. Point $x_p^n$ is actually the same node as $x_{p+1}^2$, and $x_p^{n-1}$ is the
same node as $x_{p+1}^1$.

The derivatives of  functions in the overlapped nodes are approximated by taking the
average of their values evaluated from the two intervals involved:
    \begin{eqnarray}
    \left. \D{x}{y} \right |_{x_p^{n-1}} = \frac{1}{2} \left (
    \left. \sum_{i=0}^{N} y_p^i   \D{x}{\phi_i(x)} \right |_{x_p^{n-1}} +
    \left. \sum_{i=0}^{N} y_{p+1}^i   \D{x}{ \phi_i(x)} \right |_{x_{p+1}^1}
     \right ),  \\
    \left. \D{x}{y} \right |_{x_p^n} = \frac{1}{2} \left (
    \left. \sum_{i=0}^{N} y_p^i   \D{x}{\phi_i(x)} \right |_{x_p^n} +
    \left. \sum_{i=0}^{N} y_{p+1}^i   \D{x}{ \phi_i(x)} \right |_{x_{p+1}^2}
     \right ).   \nonumber
    \end{eqnarray}
The approximation to the second order derivatives can be found using the same average for the
nodes in the overlapped region.

\subsection{Backward Differentiation Formula Method}

By employing the multi-domain decomposition methods combined with the Chebyshev collocation
methodology, the given set of partial differential equations Eq. (\ref{SMAEq}) can be
converted into a DAE system, which can be generically written in the following form:
    \begin{eqnarray}
       \vec M  \frac{d \vec X}{dt}  + \vec N(t, \vec X, g(t)) = 0,
    \end{eqnarray}
where $\vec X$ is the vector collecting all the variables we are solving for, $\vec M$ is
a singular matrix, $\vec N$ is a vector collecting nonlinear functions produced by spatial
 discretization. The resultant
DAE system is a stiff system and has to be solved by an implicit algorithm. Here the second
order backward differentiation formula method is employed for this purpose. By discretizing
the time derivative using the second order backward approximation, the DAE system can be
converted into an algebraic
 system at each time level, which can formally written as follows:
    \begin{equation}
 \vec M \left( \frac{3}{2} \vec X^n - 2 \vec X^{n - 1} + \frac{1}{2}
 \vec X^{n - 2} \right) + \Delta t \vec N \left( t_n, \vec X^n, g(t_n) \right) = 0 ,
    \end{equation}
where $n$ denotes the current computational time layer. For each computational
 time layer, iterations must be carried out using
Newton's method for $\bf X^n$ by use of $\bf X^{n-1}$ and $\bf X^{n-2}$. Starting from the
initial value, the vector of unknowns ${\bf X}$ can be solved for at all specified time
instances employing this algorithm.


\section{Numerical Experiments}

A series of numerical experiments have been carried out  to
investigate the nonlinear wave propagations in the SMA rod involving
phase transformations.  All experiments reported here have been
performed on a $\textrm{Au}_{23}\textrm{Cu}_{30} \textrm{Zn}_{47}$
rod, with a length of $1$ cm.  The physical parameters, except $\nu$
and $k_g$, for this specific material are taken the same as those in
\cite{Niezgodka1991}, which are listed as follows for convenience:
\begin{eqnarray*}
 k_{1}=480\, g/ms^{2}cmK, \quad  k_{2}=6\times10^{6}g/ms^{2}cmK,\qquad
  k_{3}=4.5\times10^{8}g/ms^{2}cmK,   \\
\theta_{1}=208K,\quad  \rho=11.1g/cm{}^{3},\quad
c_{v}=3.1274g/ms^{2}cmK, \quad k=1.9\times10^{-2}cmg/ms^{3}K.
\end{eqnarray*}

Experiments indicate that the Ginzburg coefficient $k_g$  should be relatively
small compared to $k_1$, so it is taken as $k_g =10 g/ms^2$ first, by referring
to Ref.\cite{Bubner1996}. The internal
 friction coefficient is not an easily measurable quantity. In what follows, we assume
 it to be a small  fraction (2\%) of $k_1$, that is approximately $10 g/(cm)(ms))$.
The entire rod is divided into $10$ sub-intervals, in each interval there are $15$ nodes used
 for spatial discretization. All simulations have been carried out for the time span $[0,0.1]$ ms,
and the time step-size for the integration is chosen as $2.5 \times 10^{-5}$ms.

The first numerical experiment for nonlinear wave propagations in
the SMA rod is performed with a higher temperature $ \theta=310$ K,
 for which there is no phase transformation.
 Other initial conditions are chosen  $u=v=s=0$, and  mechanical loading
is employed in terms of a stress impact at right end, as follows:
    \begin{eqnarray}
         g(t) =\begin{cases}
            4\times 10^3, &   0 \leq t  \leq 0.005  \cr   0    , &  t > 0.005
             \end{cases}
    \end{eqnarray}
which can be regarded as an approximation to a pulse stress impact on the SMA rod.

\begin{figure}   \begin{center}
    \includegraphics[width=7.6cm, height=6cm]{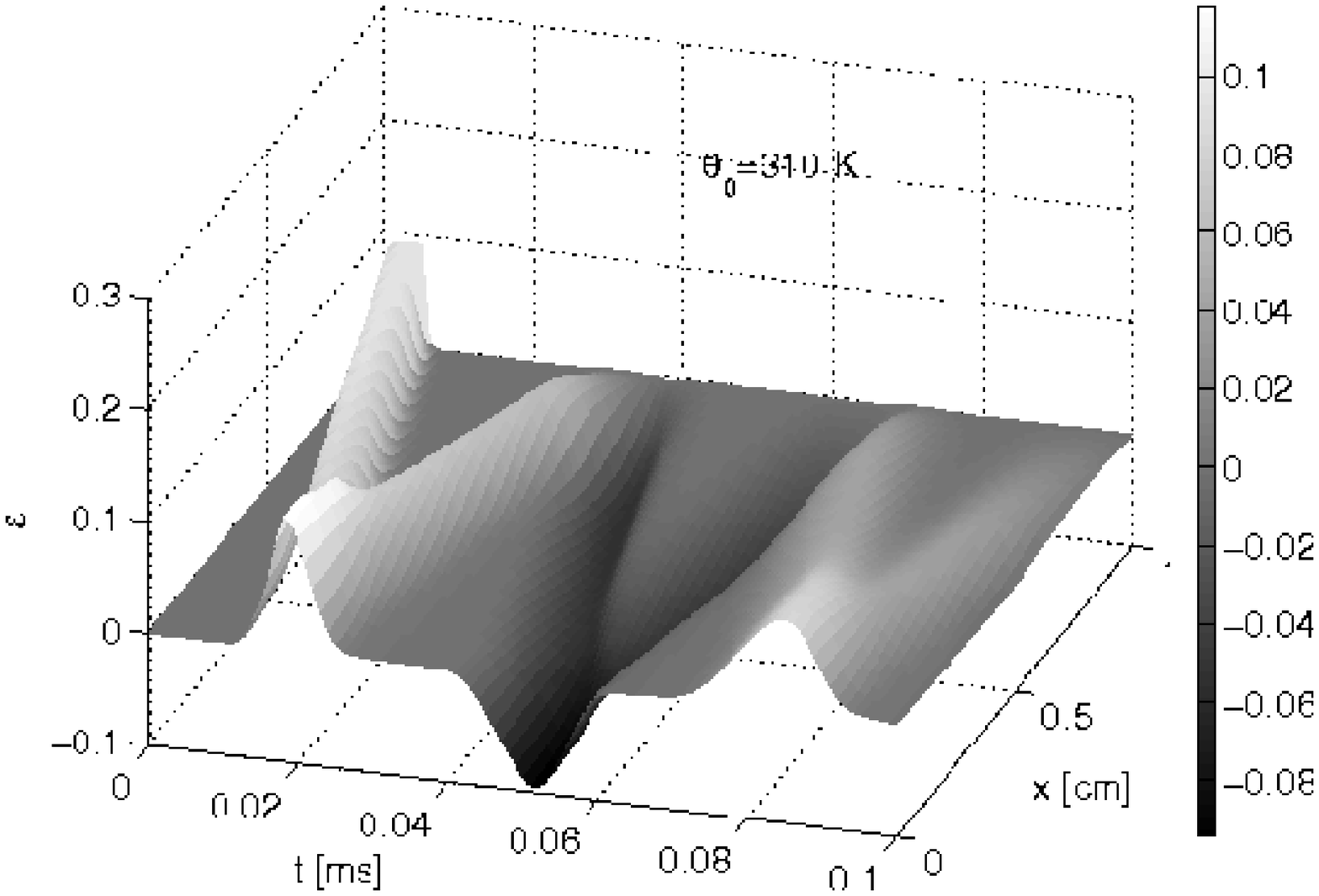}
    \includegraphics[width=7.6cm, height=6cm]{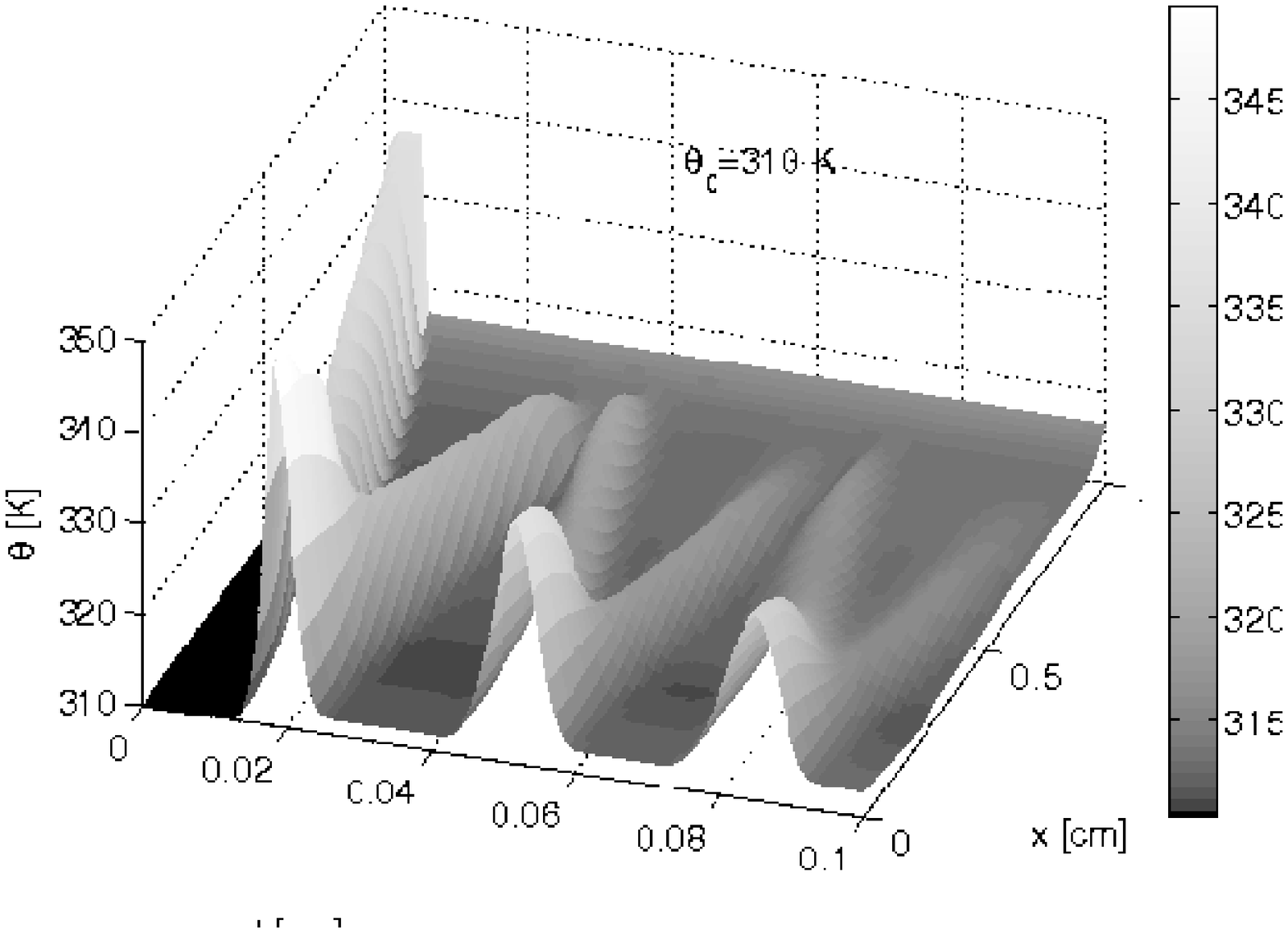} \\
    \includegraphics[width=8cm, height=6cm]{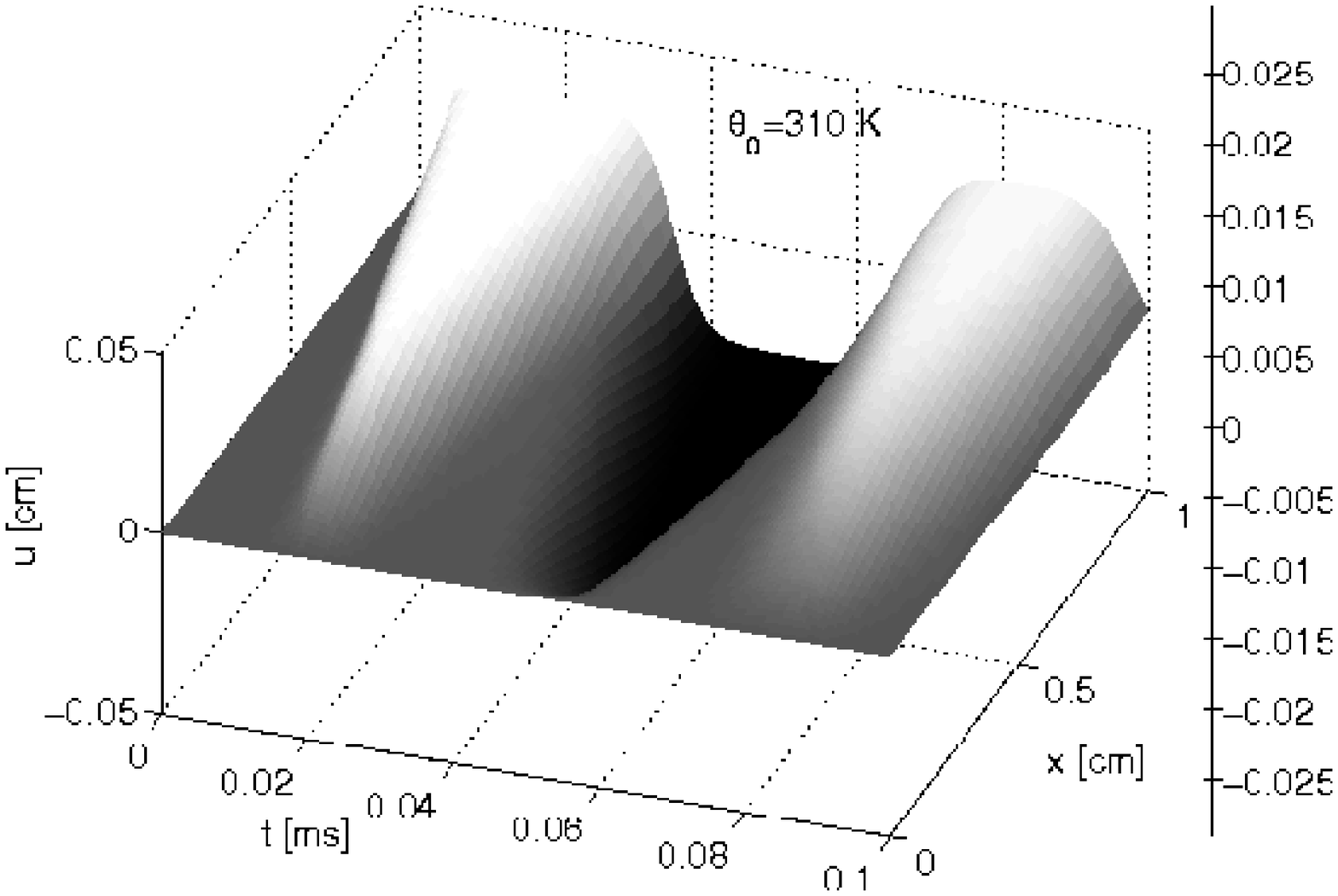}
    \includegraphics[width=6.5cm, height=5.5cm]{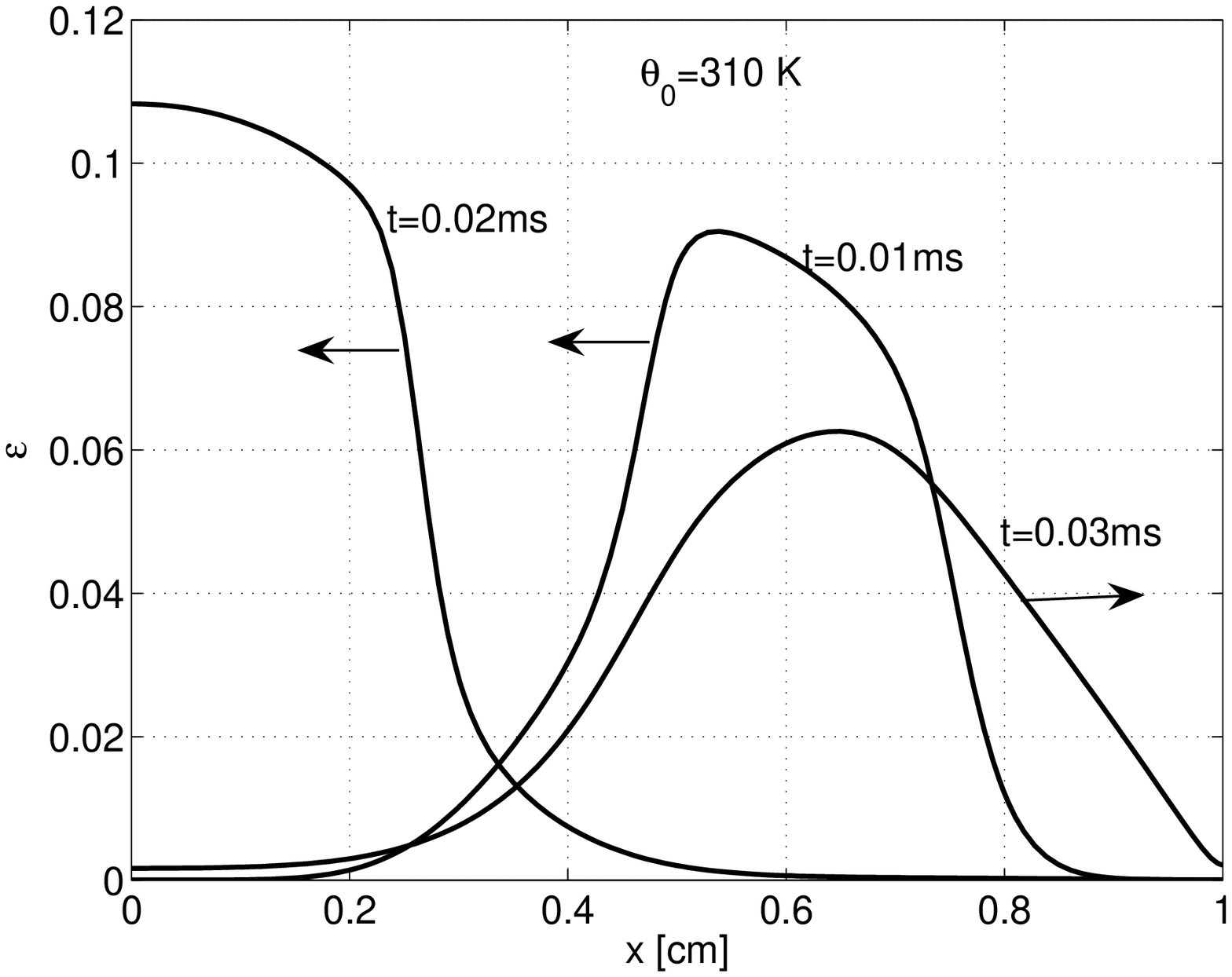}
     \caption{Nonlinear thermo-mechanical wave propagations in a shape memory alloy
      rod caused by a stress impact, initial temperature is $\theta=310$ K. }
     \label{Pulse310}
      \end{center}  \end{figure}

The mechanical wave propagations are presented by the strain evolution
while thermal waves by temperature evolution in Figure (\ref{Pulse310}).
It is shown that the impact induced waves start from $x=1$ and propagate
along the negative $x$ direction,  hit the opposite boundary at $x=0$ and are bounced back.
The temperature evolution indicates that there are associated thermal waves induced
by the stress impact loading, due to the thermo-mechanical coupling effects.
 The propagation patterns of the thermal waves
are similar to those of mechanical waves.  The evolution of the
 displacement distribution due to the stress impact is also presented
in Figure (\ref{Pulse310}), in the left bottom sub-figure. To
clarify the patterns of wave propagations,  the strain distributions
in the SMA rod  at three chosen time instants are plotted in the
right bottom sub-figure in Figure (\ref{Pulse310})
($t=0.01,~0.02,~0.03$ ms, respectively). The arrow attached to each
wave profile is to indicate its propagation direction.  It is seen
that the strain distributions in the SMA rod are always smooth and
no obvious sharp jump occurs, since only austenite is stable with
this temperature and there is no phase transformation. The amplitude
of the wave decreases and the wave peak becomes broader during the
propagation, it can be easily explained by the fact that dissipation
effects and dispersion effects, caused by internal
 friction, capillary effects, and thermo-mechanical coupling, are all included in the model.
 The average wave propagation
 speed can be estimated by the location of the wave  frontier or wave peak plotted in the
 figure. With the
current  initial temperature, the strain wave is bounced back from $x=0$ and its peak
is located around $x=0.65$ cm when $t=0.003$ ms. This experiment simulates the
nonlinear thermo-mechanical waves like those in regular thermo-elastic materials.

\begin{figure}  \begin{center}
    \includegraphics[width=7.6cm, height=6cm]{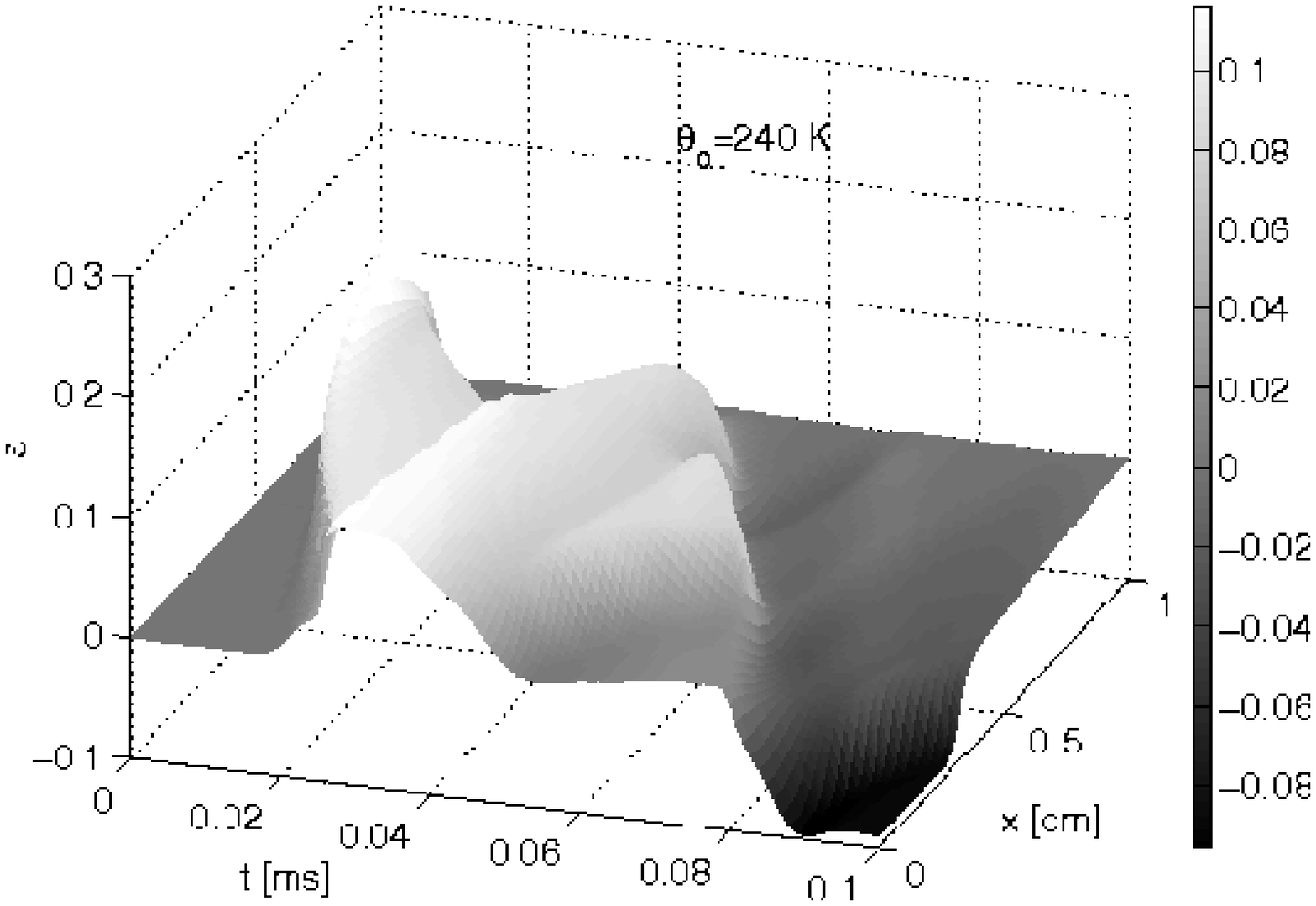}
    \includegraphics[width=7.6cm, height=6cm]{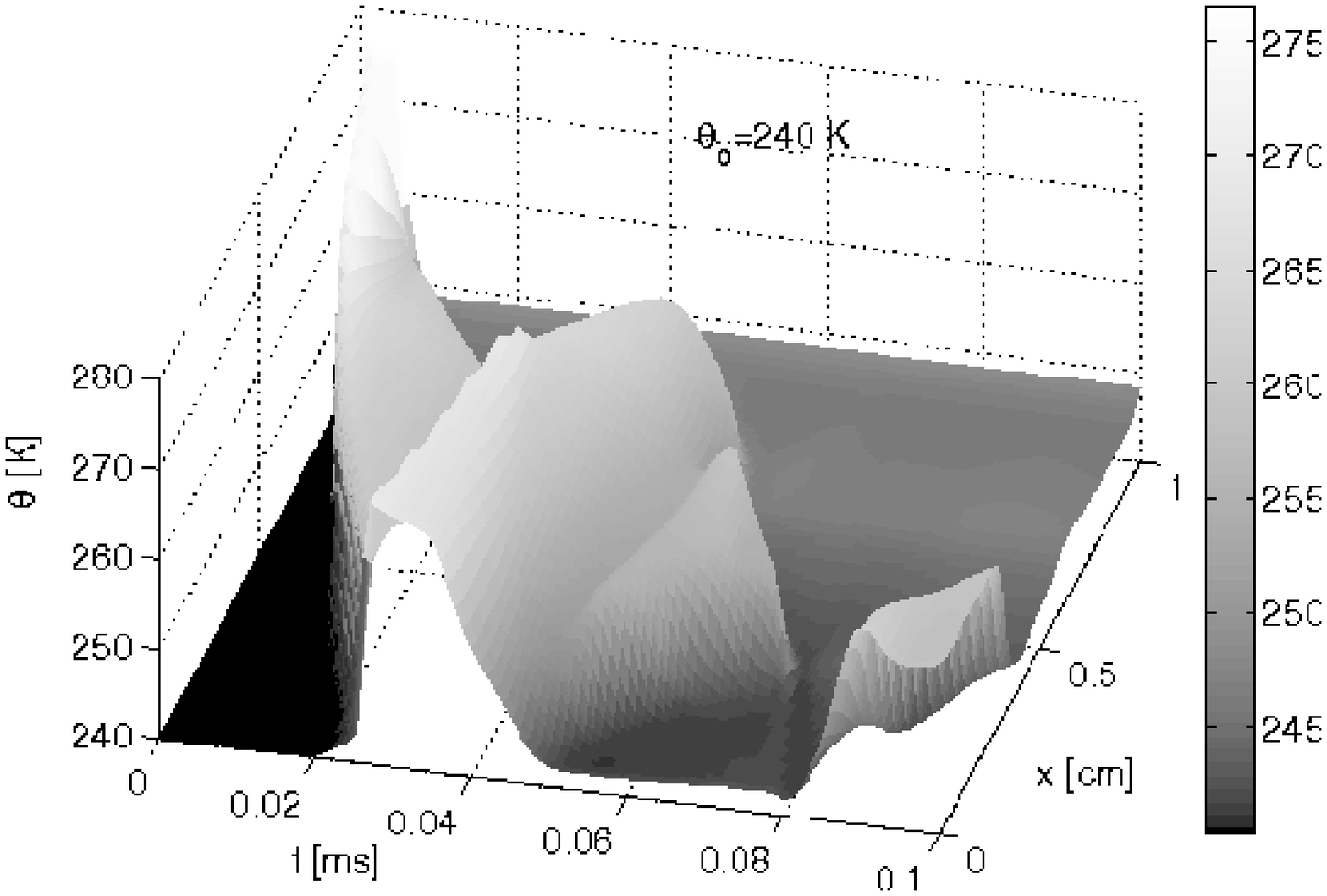} \\
    \includegraphics[width=8cm, height=6cm]{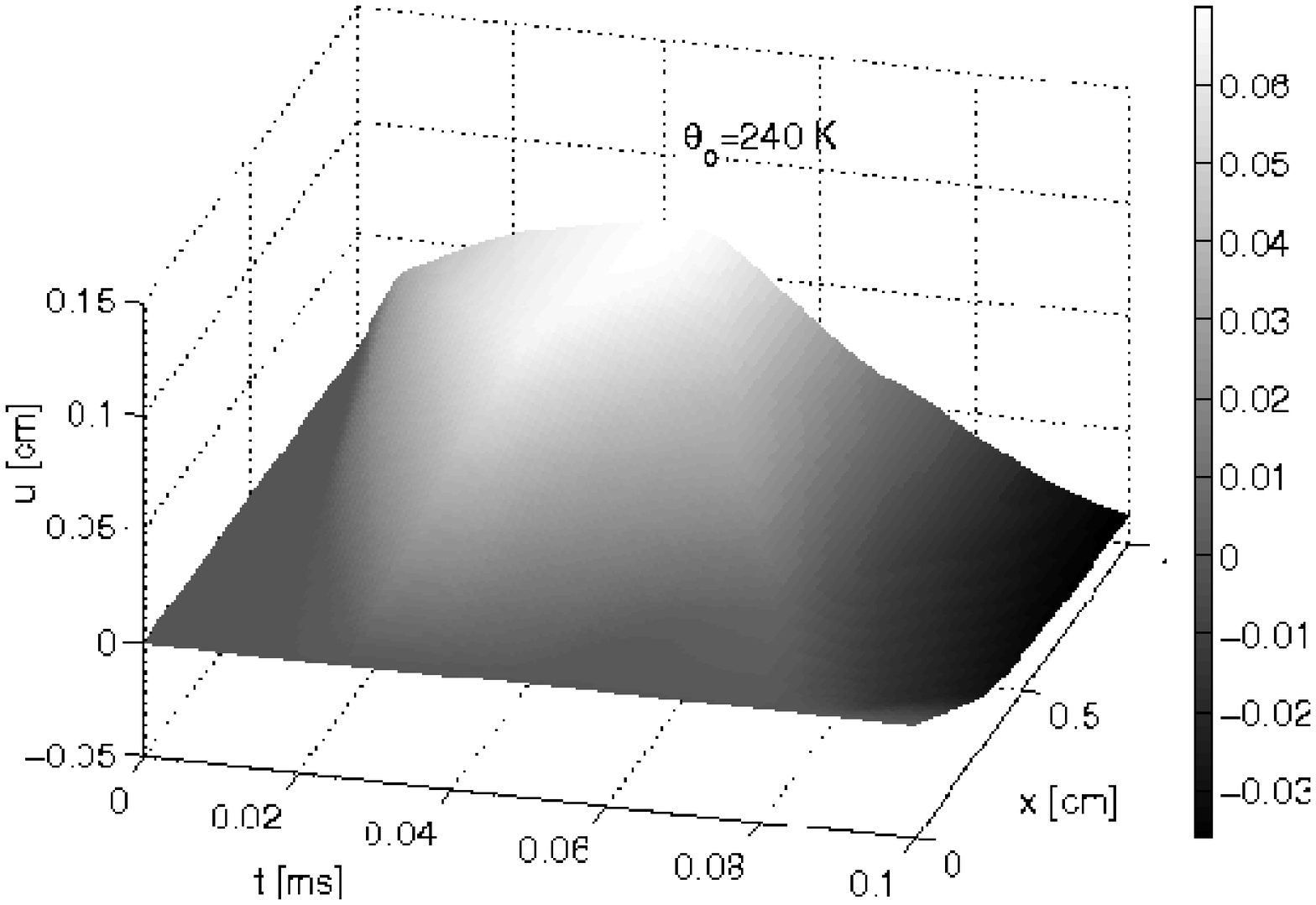}
    \includegraphics[width=6cm, height=6cm]{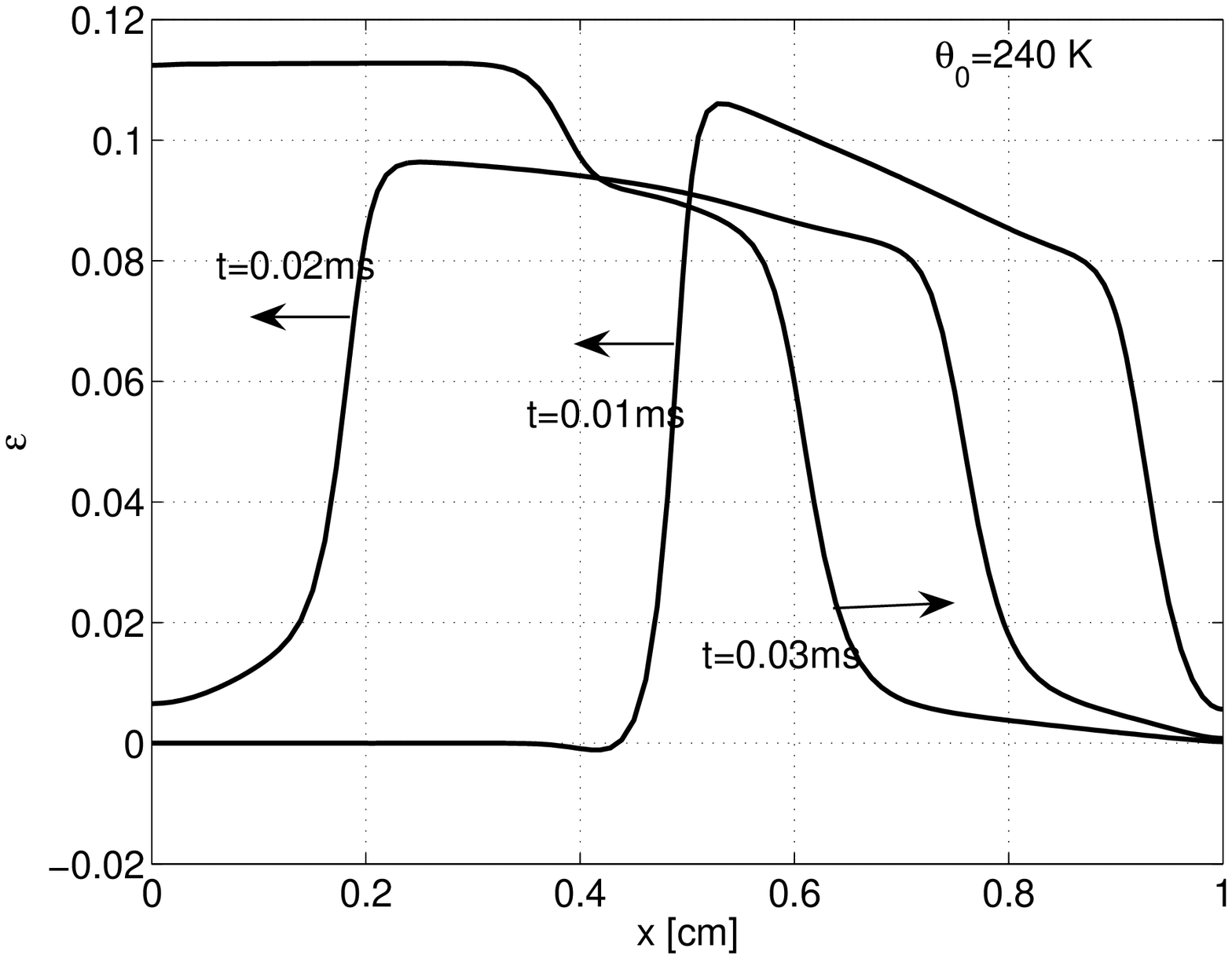} \\
    \caption{Nonlinear thermo-mechanical wave propagations involving phase transformations
     in a shape memory alloy rod caused by a stress impact, initial temperature is
     $\theta=240$ K.}
     \label{Pulse240}
      \end{center}  \end{figure}

The second example deals with the same computational conditions and
loading, except that the initial temperature now is set
$\theta=240K$, for which both martensite and austenite may co-exist
in the SMA rod. The numerical results for this case are presented
similarly in Figure (\ref{Pulse240}). It is easy to see that the
strain and temperature waves are not as regular as those in the
first experiment. There are some plateaus clearly shown in the
strain and temperature figures, which can be related to martensite
and austenite in the SMA rod. The frontier of the waves are now more
easily identified since abrupt jumps occur in the strain
distribution, which are caused by phase transformations between
austenite and martensite. In the displacement evolution now we have
only one peak within the simulated time span, while two peaks are
found
 in the first experiment.  On the three chosen time instants,  the strain distributions are not as
smooth as those at high temperature. At $t=0.03$ ms, the wave
frontier is around $x=0.65$ cm, propagating along the positive $x$ direction. It indicates that
the wave speed is a little lower than that in the first experiment, as indicated by the analysis
given in section 3.  Similarly, there are thermal waves caused by the mechanical loading
 due to the coupling effects.

\begin{figure}    \begin{center}
    \includegraphics[width=7.6cm, height=6cm]{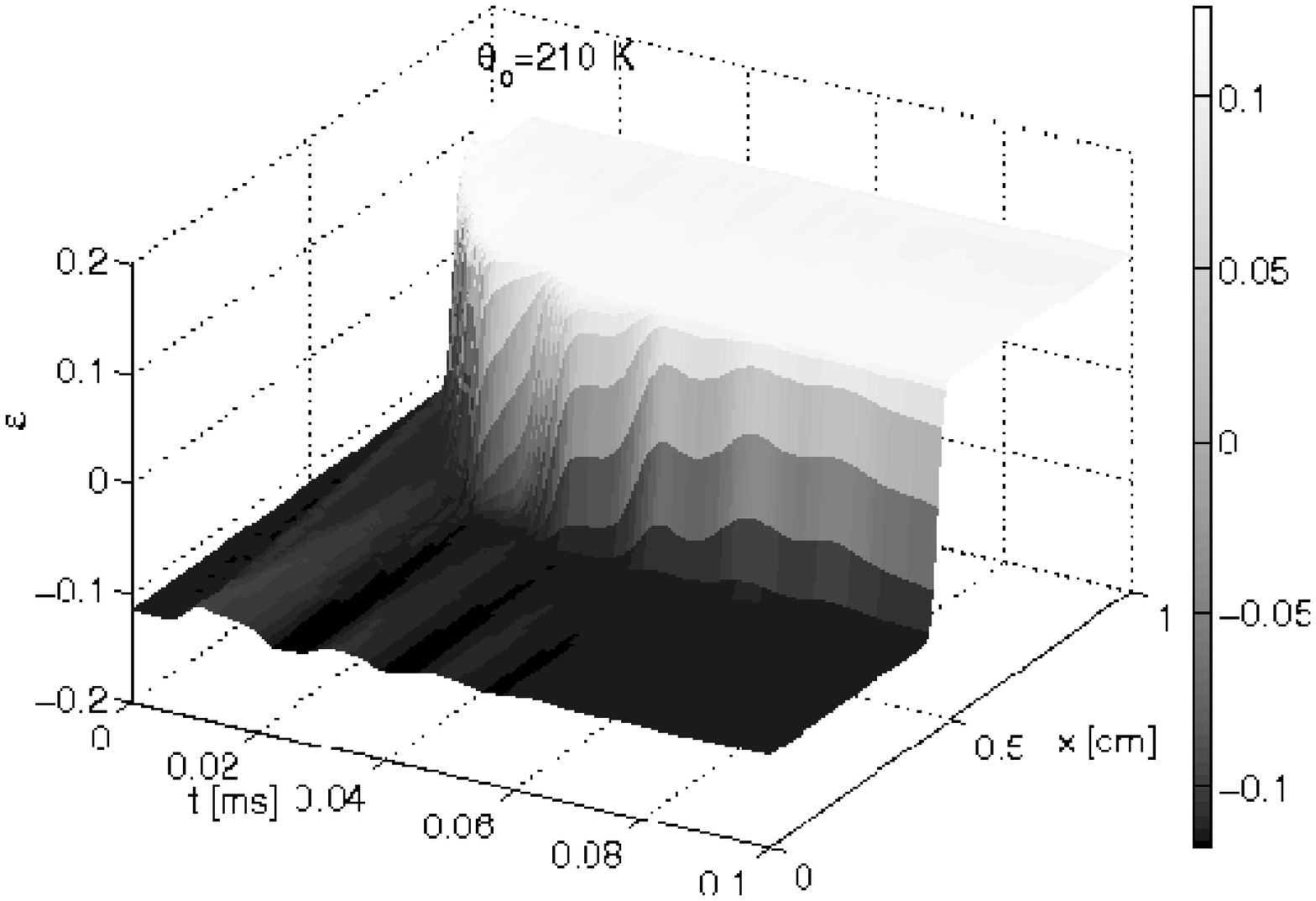}
    \includegraphics[width=7.6cm, height=6cm]{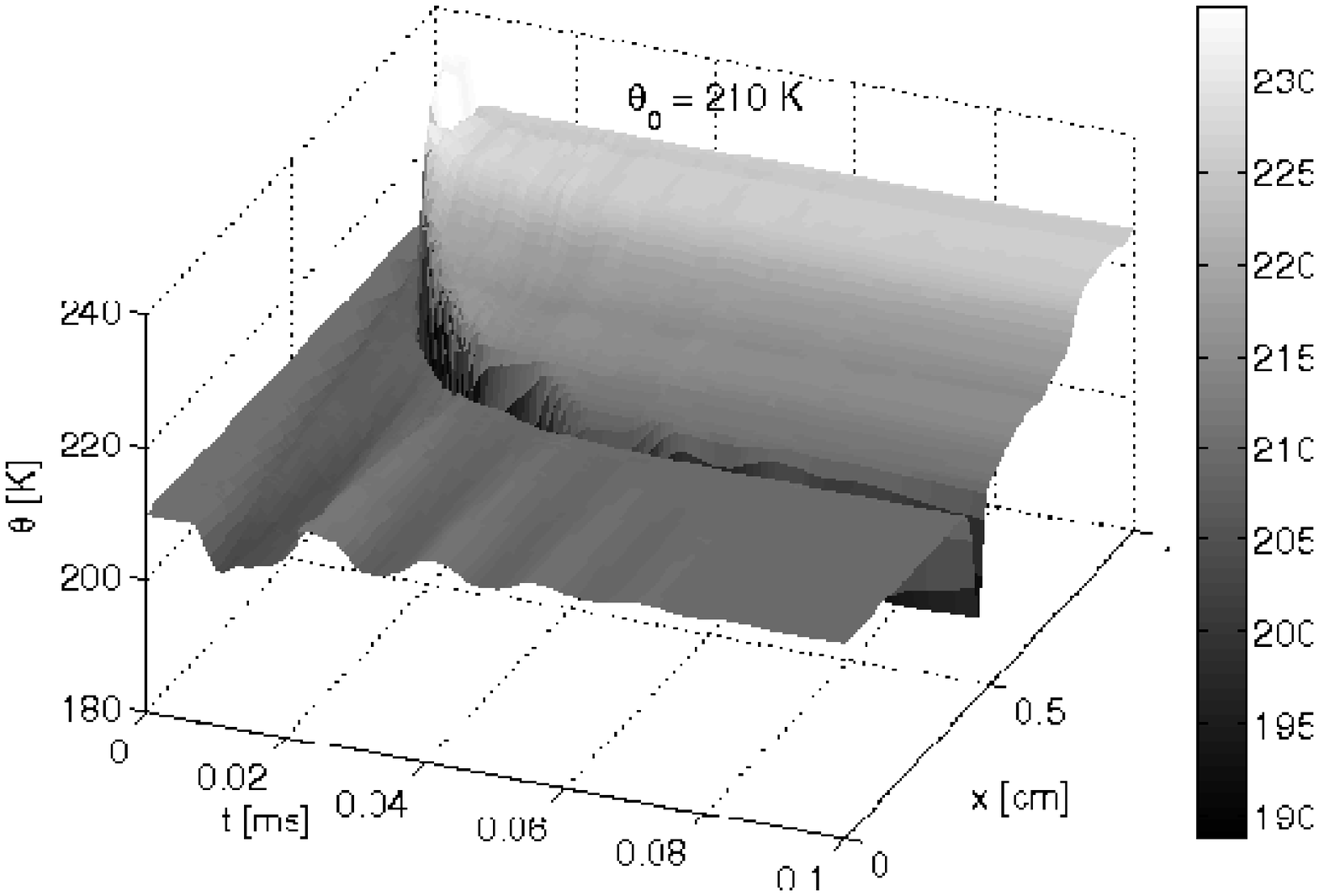} \\
    \includegraphics[width=8cm, height=6cm]{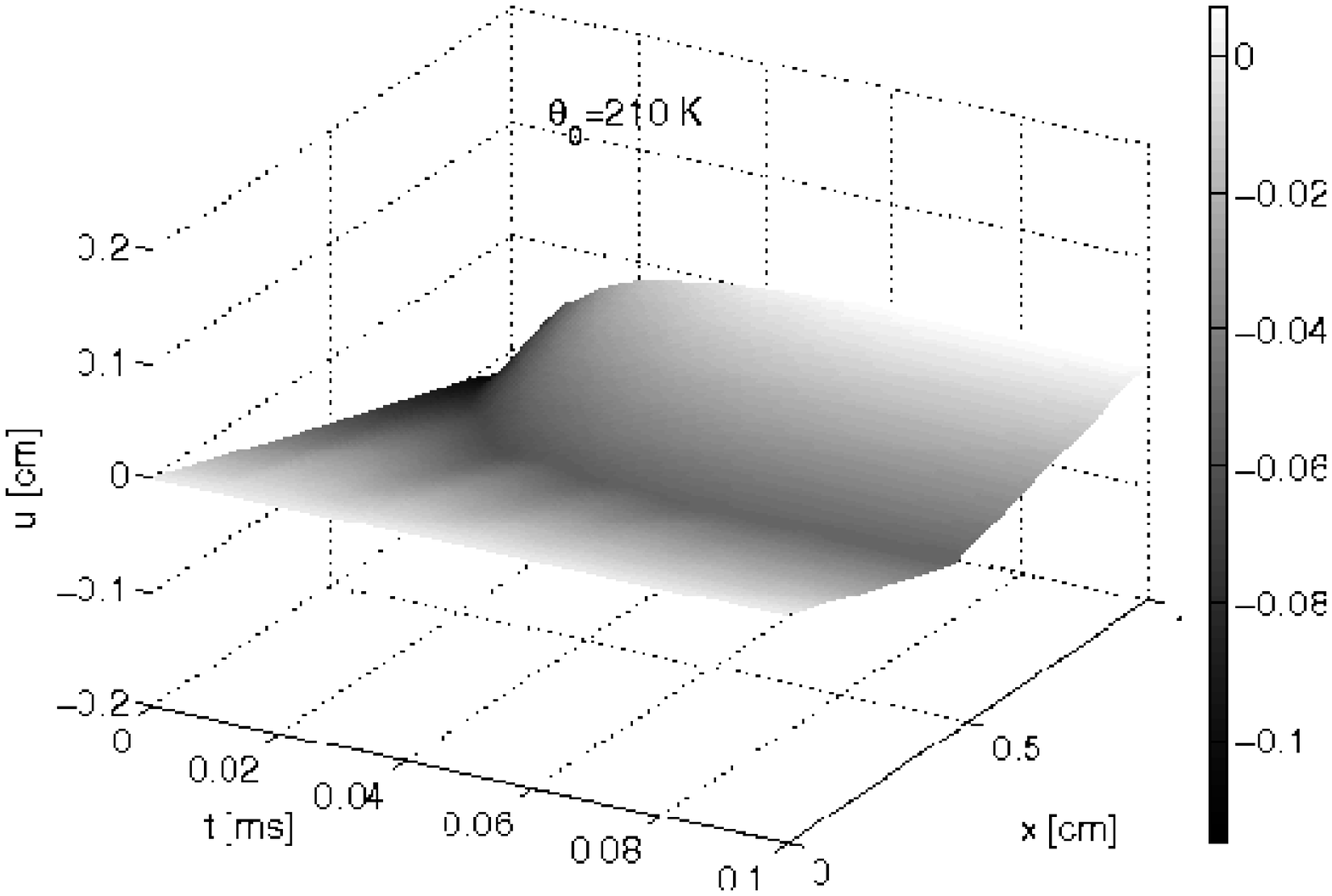}
    \includegraphics[width=6cm, height=5cm]{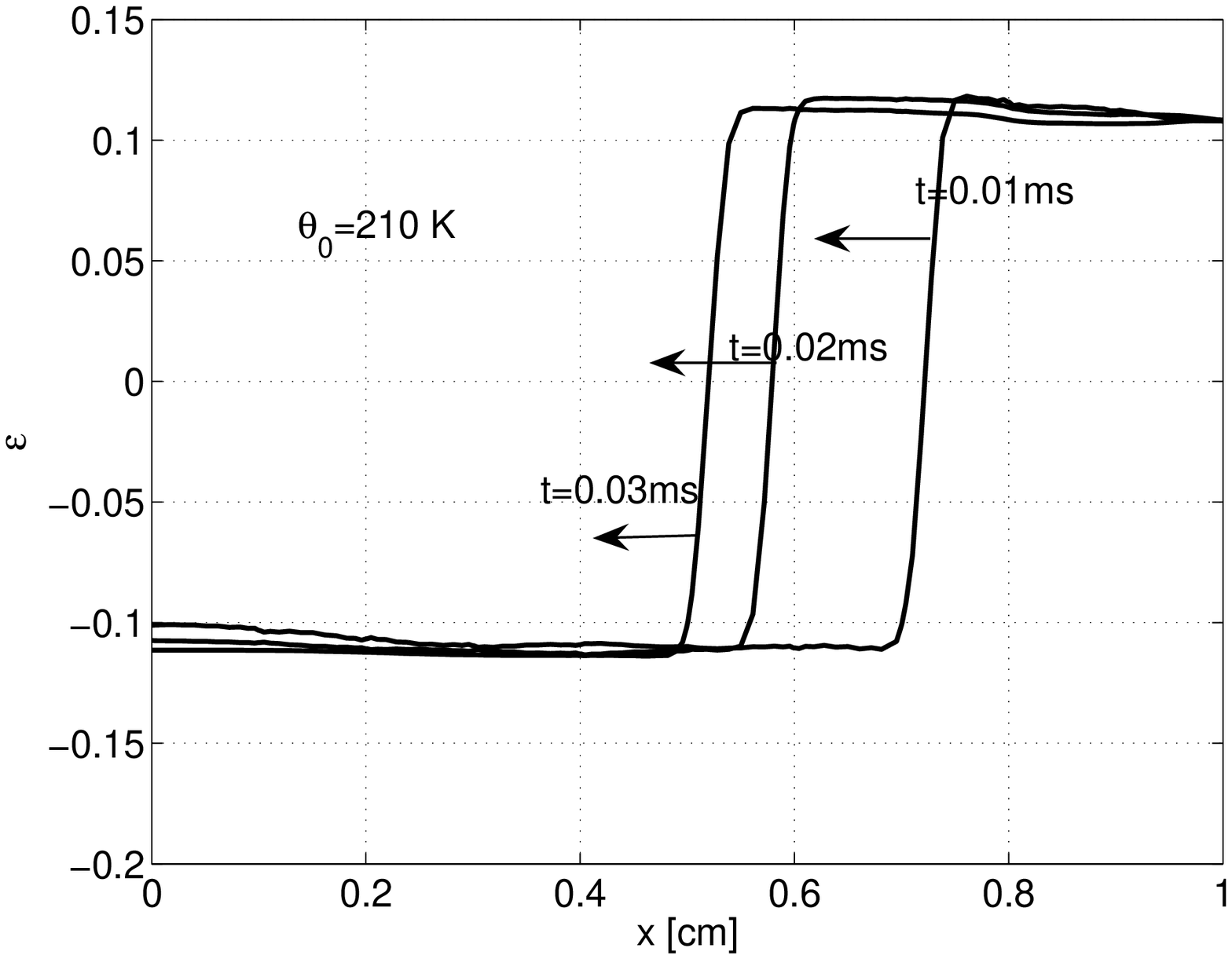}
    \caption{Nonlinear thermo-mechanical wave propagation in the shape memory alloy rod
     caused by a stress impact, involving martensite phase transformations.
     Initial temperature is $\theta=210 K$}
      \label{Pulse210}
      \end{center}  \end{figure}

For the third experiment,  the initial temperature is set at
$\theta=210$ K. Because only martensite is stable at this
temperature, the initial condition is chosen such that  the SMA rod
is originally at martensite minus state, for which the displacement
is set $u=-0.115x$ so $\eps_0=-0.115$. This strain value is one of
the local minima of the non-convex potential
 energy plotted in Figure~(\ref{GinzburgLandau}), and calculated in Equation~(\ref{EquiPoint}).
 Numerical results for this case are presented in
Figure (\ref{Pulse210}). It is seen that the entire SMA rod is
divided into two domains, one consists of martensite plus (with
$\eps \approx 0.115$), and the other one - martensite minus ($\eps
\approx -0.115$). The interface between the two domains is driven by
the impact stress loading, as sketched in the strain evolution plot
and the wave profiles on the chosen time instants in Figure
(\ref{Pulse210}). With the given computational conditions, the SMA
rod is converted from martensite minus to plus from the end which
under impact loading, the phase boundary is driven to propagate
along the negative $x$ direction. Due to the hysteretic nature of
the phase transformation, the input mechanical energy is dissipated
continuously during the propagation of the phase boundary, and is
not able to convert the whole rod into martensite plus. The
interface stops at around $x=0.5$. Correspondingly, there are also
thermal waves accompanying the martensite transformation.
 The wave propagation speed is much lower compared to those in the previous
experiments, and the wave speed changes more remarkably during the
propagation process, as indicated by the plot of wave profiles plot
 at chosen time instants.
 At $t=0.03$ ms, the wave frontier is at $x=0.5$ cm and is unable to move further toward the
 end $x=0$.

\begin{figure}  \begin{center}
    \includegraphics[width=8cm, height=6cm]{Strain310.eps}
    \includegraphics[width=6cm, height=5cm]{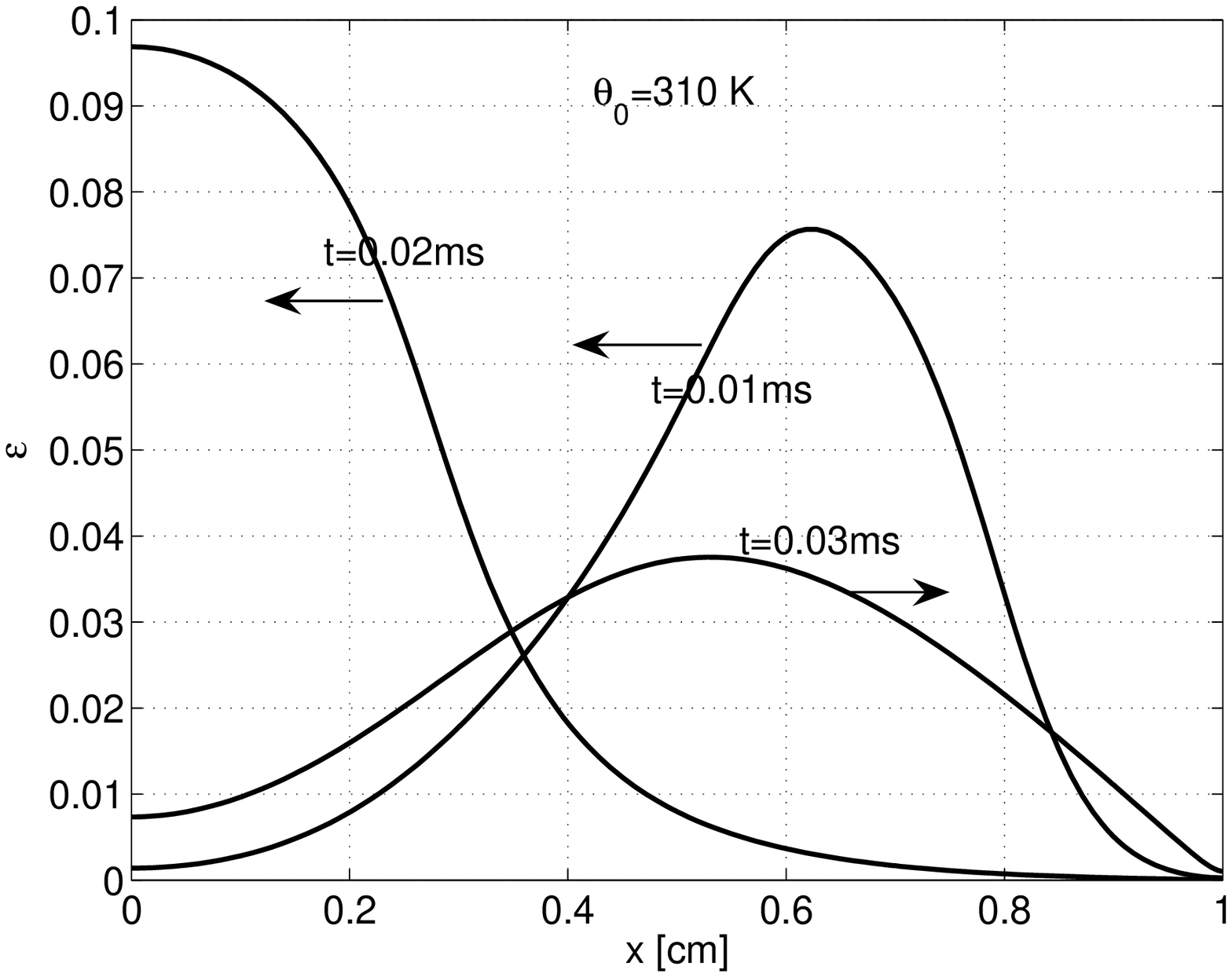}  \\
     \caption{Numerical analysis of the dissipation effect of internal friction on
  wave propagations in a shape memory alloy rod.
 Initial temperature is $\theta=310$ K, $\nu =30$.}
      \label{Pulse310Vis}
      \end{center}  \end{figure}

The forth experiment is to investigate the dissipation effects due
to internal friction in the material. As analyzed in section 3, the
dissipation effects are independent of temperature, so we set the
initial temperature at $\theta_0=310$ K to exclude phase
transformation, so that the comparison will be easier.  The internal
friction $\nu$ is set three times larger at $30$, and all other
computational parameters are chosen the same as those in the first
experiment.  The strain evolution and wave profiles at the same
three chosen time instants are presented in
Figure~(\ref{Pulse310Vis}). By comparing with those in
Figure~(\ref{Pulse310}), it is observed that the dissipation effects
are enhanced, the peak values of the wave decrease faster. At
$t=0.01$ ms, the peak value of the wave profile is around $0.075$,
located around $x=0.6$ cm, while the counterparts with $\nu=10$ are
peak value $0.089$ at around $x=0.055$ cm.  This indicates that when
$\nu$  is increased, not only the peak value dissipated faster, but
the wave speed was also slightly slowed down.

\begin{figure}  \begin{center}
    \includegraphics[width=8cm, height=6cm]{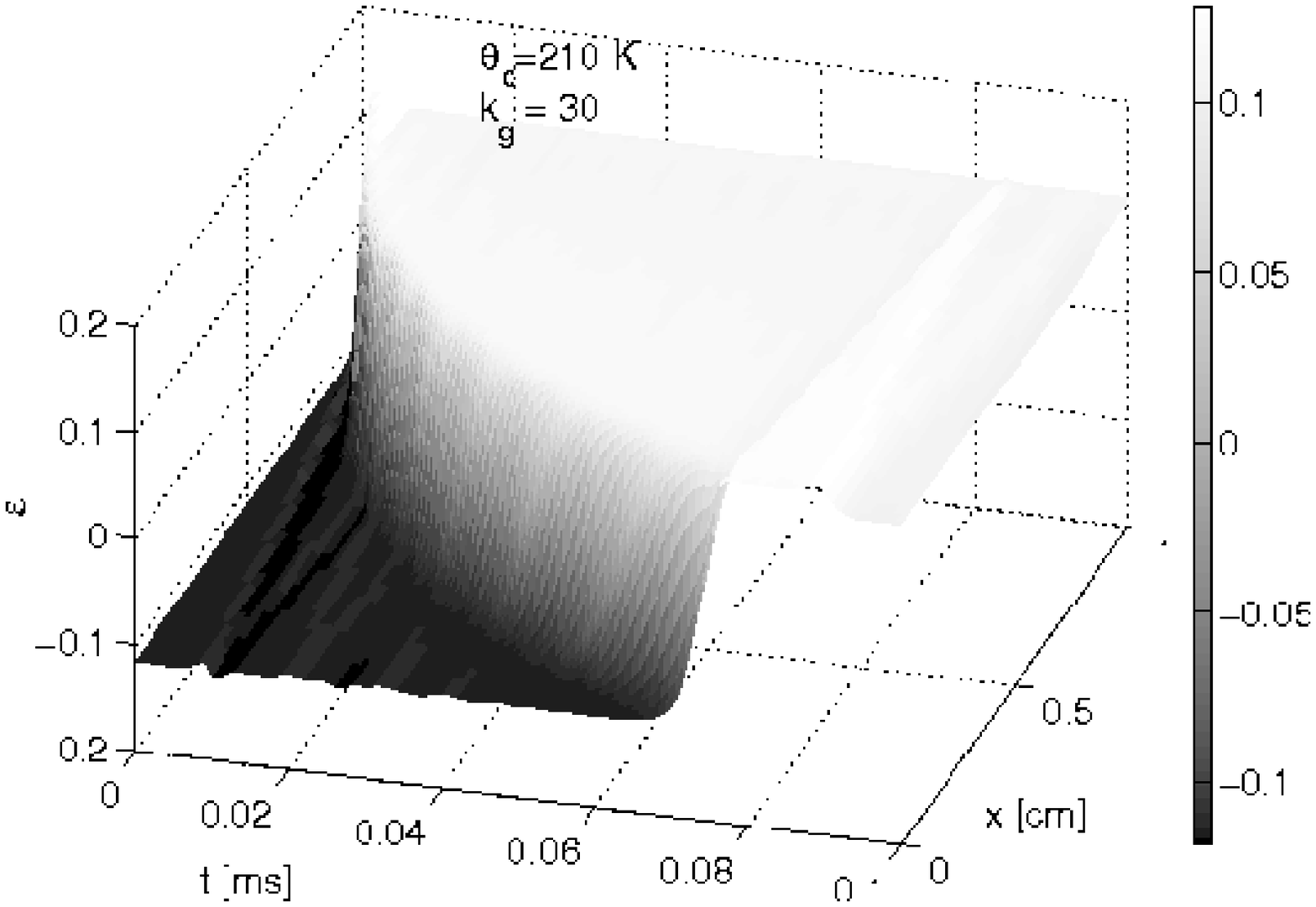}
    \includegraphics[width=6cm, height=5cm]{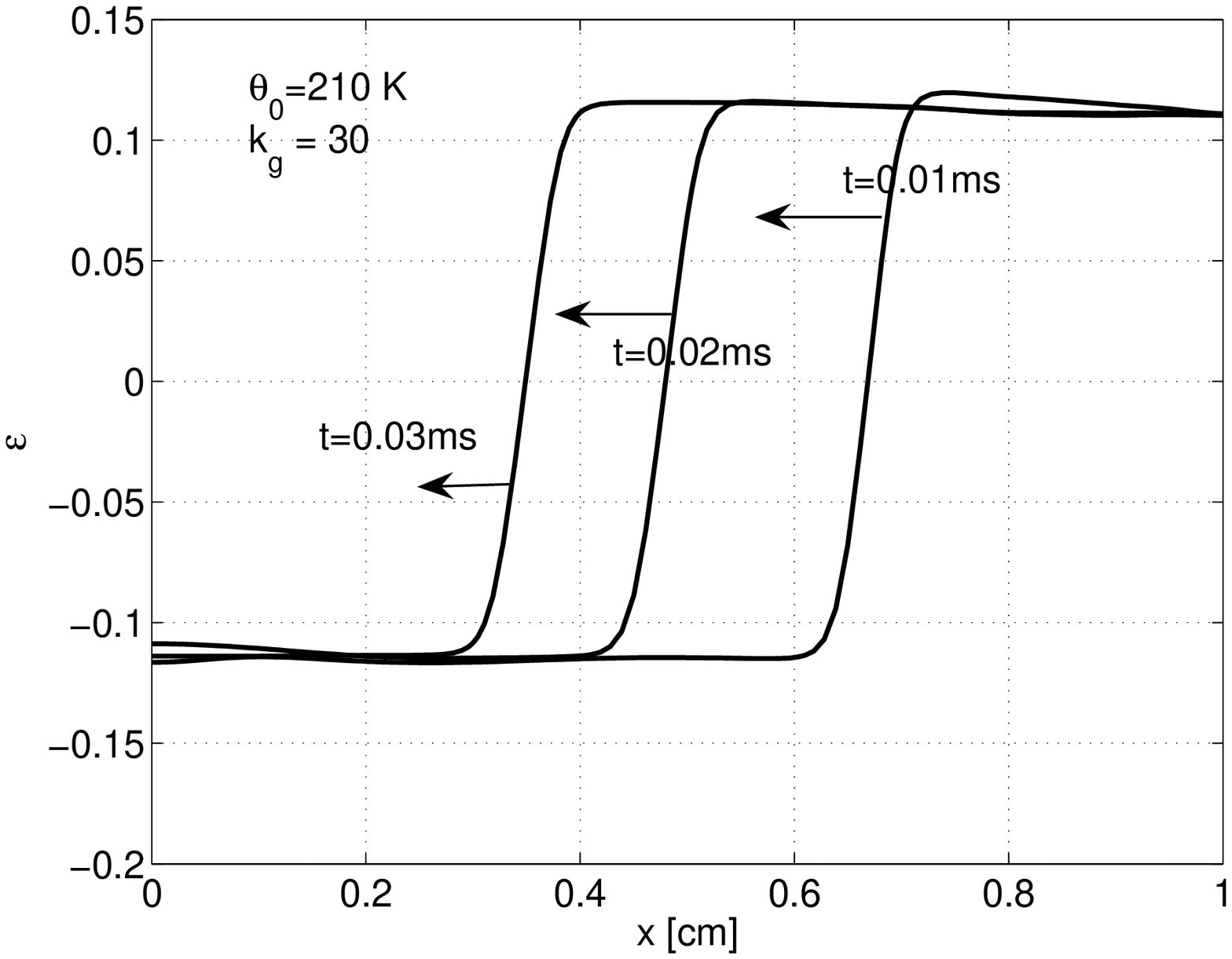} \\
     \caption{Numerical analysis of the dispersion effects caused by capillary effects
     in  wave propagations in a shape memory alloy rod.
 Initial temperature is $\theta=210$ K, $k_g =30$.}
      \label{Pulse210Kg}
      \end{center}  \end{figure}

The final experiment is to show numerically the dispersion effects
due to the Ginzburg term in the wave equation.  As indicated by
Equation~(\ref{WaveDisper}), the dispersion effects are more
pronounced at low temperature since $k_L=k_1(\theta-\theta_1)$ will
be smaller.  To perform the analysis, the initial conditions are set
the same with those in the third experiment, except that $k_g$ is
set three times larger at $30$. The strain evolution and wave
profiles for three chosen time instants are presented in
Figure~(\ref{Pulse210Kg}).  By comparing the results with those in
Figure~(\ref{Pulse210}), it can be seen that the entire rod is still
divided into two domains, one for martensite plus ($\eps \approx
0.115$) and the other - for martensite minus ($\eps \approx
-0.115$). However, the wave propagation speed is faster with larger
$k_g$ value. The interface between martensite minus and plus is
located at around $x=0.35$ cm when $t=0.03$ ms, while for $k_g=10$
 it is at $x=0.5$ cm. This observation agrees with the linearization
analysis carried out in section 3.  At the same time, the entire rod
is converted from martensite minus to plus, which indicates that
smaller amount of input energy is demanded for the phase
transformation when $k_g$ value is larger. In other words, the phase
transformation becomes easier to deal with when the capillary
effects are enhanced.

From the above numerical experiments follow that nonlinear
thermo-mechanical wave propagations caused by impact loadings in the
SMA rod can be remarkably influenced by the material temperature,
internal friction, and capillary effects.  Thermal waves could be
induced by impact mechanical loadings. Wave propagation patterns are
more complicated when phase transformations
 are involved, and the dynamic response of the material in this case
 is very different from those with no phase transformations.


\section{CONCLUSIONS}

In this paper,  a mathematical model for the analysis of wave propagations in a shape memory
 alloy rod induced by a stress impact was constructed. The modified
 Ginzburg-Landau-Devonshire
theory was employed for modelling dynamic processes in SMA rods. The
first order martensite phase transformations and thermo-mechanical
coupling were incorporated into the model. A multi-domain
decomposition method
 was employed in conjunction with the Chebyshev collocation method
for spatial discretization, and the backward differentiation formula was used for solving the
resulting differential-algebraic system. The nonlinear thermo-mechanical wave propagations
 in the SMA rod were simulated with various initial temperatures (with and without phase
 transformation),   the effects of phase transformations on the wave propagations were
 analysed numerically, along with the effects of  internal friction and capillary.



\newpage

\newpage

     \begin{figure}    \begin{center}
     \includegraphics[scale=0.6]{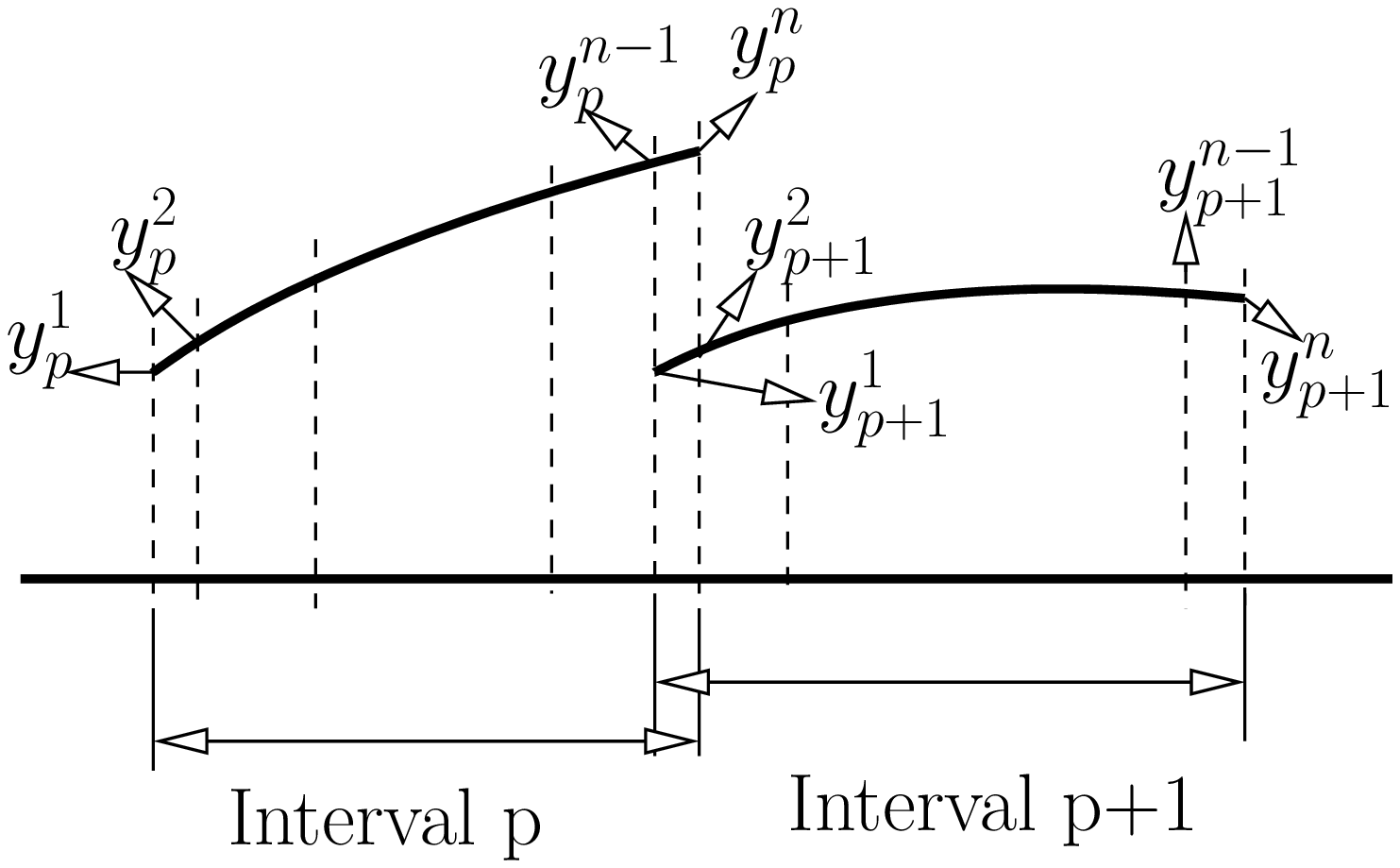}
     \caption{Mechanical dissipation due to martensite phase transformation }
     \label{MechDissip}
      \end{center}  \end{figure}

\newpage

\newpage

\newpage

\newpage

\end{document}